\documentclass[useAMS,usenatbib]{mn2e}

\usepackage{lscape}
\usepackage{graphicx}
\usepackage{natbib}
\usepackage{graphics}

\title[Morphological Fractions in WINGS]{Morphological Fractions of
  Galaxies in WINGS Clusters:\\revisiting the Morphology-Density Paradigm}
\author[G. Fasano et al.]{
\parbox[t]{\textwidth}{
G. Fasano$^{1}$\thanks{E-mail:giovanni.fasano@oapd.inaf.it}, 
B.M. Poggianti$^{1}$, 
D. Bettoni$^{1}$, 
M. D'Onofrio$^{2}$, 
A. Dressler$^{3}$, 
B. Vulcani$^{4}$,
A. Moretti$^{1}$,
M. Gullieuszik$^{1}$,
J. Fritz$^{5,6}$,
A. Omizzolo$^{1,7}$,
A. Cava$^{8}$,
W.J. Couch$^{9}$,
M. Ramella$^{10}$ and
A. Biviano$^{10}$.}\\
\\
$^{1}$INAF, Osservatorio Astronomico di Padova, Vicolo Osservatorio 5, 35122 Padova, Italy\\
$^{2}$Dip. di Fisica e Astronomia G. Galilei, Universit\`a di Padova, Vicolo dell'Osservatorio 2, 35122 Padova, Italy\\
$^{3}$The Observatories of the Carnegie Institution of Washington, Pasadena, USA\\
$^{4}$Kavli Institute for the Physics and Mathematics of the Universe (WPI),\\ Todai Institutes for Advanced Study, the University of Tokyo, Kashiwa, 277-8582, Japan\\
$^{5}$Centro de Radioastronom\`ia y Astrof\`isica, CRyA, UNAM,\\ Campus Morelia, A.P. 3-72, C.P. 58089, Michoac\`an, Mexico\\
$^{6}$Sterrenkundig Observatorium, Universiteit Gent, Krijgslaan 281 S9, B-9000 Gent, Belgium\\
$^{7}$Specola Vaticana, 00120 Stato Citt\`a del Vaticano\\
$^{8}$Observatoire de Gen\`eve, Universit\`e de Gen\`eve, 51 Ch. des Maillettes, 1290 Versoix, Switzerland\\
$^{9}$Center for Astrophysics, Swinburne University of Technology, Australia\\
$^{10}$INAF, Osservatorio Astronomico di Trieste, Via G.B. Tiepolo, 11, 34131 Trieste, Italy}

\begin{document}

\date{Accepted .... Received .....; in original form .....}

\pagerange{\pageref{firstpage}--\pageref{lastpage}} \pubyear{2012}

\maketitle

\label{firstpage}

\begin{abstract}

  We present the Morphology-Density and Morphology-Radius relations
  ($T$-$\Sigma$ and $T$-$R$, respectively) obtained from the WINGS
  database of galaxies in nearby clusters. Aiming to achieve the best
  statistics, we exploit the whole sample of galaxies brighter than
  M$_V=-$19.5 (5,504 objects), stacking up the 76 clusters of the
  WINGS survey altogether.

  Using this global cluster sample, we find that the $T$-$\Sigma$
  relation holds only in the inner cluster regions
  ($R<1/3\times R_{200}$), while the $T$-$R$ relation keeps almost
  unchanged over the whole range of local density. A couple of tests
  and two sets of numerical simulations support the robustness of
  these results against the effects of the limited cluster area
  coverage of the WINGS imaging. The above mentioned results hold for
  all cluster masses (X-ray luminosity and velocity dispersion) and
  all galaxy stellar masses (M$_*$). The strength of the $T$-$\Sigma$
  relation (where present) increases with increasing M$_*$, while this
  effect is not found for the $T$-$R$ relation.

  Noticeably, the absence/presence of subclustering determines the
  presence/absence of the $T$-$\Sigma$ relation outside the inner
  cluster regions, leading us to the general conclusion that the link
  between morphology and local density is preserved just in
  dynamically evolved regions. We hypothesize that some mechanism of
  morphological broadening/redistribution operates in the
  intermediate/outer regions of substructured (``non relaxed'')
  clusters, producing a strong weakening of the $T$-$\Sigma$ relation.

\end{abstract}


\section{Introduction} \label{secintro}

The detection of significant differences in the galaxy populations
between clusters and field dates back to \citet{HH31}, who noted that
``the predominance of early types is a conspicuous feature of clusters
in general''. Twenty years later, the prevalence in clusters of
stellar population II galaxies (early types, mainly S0s) was
interpreted by \citet{SB51} as due to galaxy collisions in the dense
cluster environment. Attempts to classify clusters according to the
galaxy morphological composition were subsequently made by
\citet{morg61} and \citet{oeml74}.

Yet, the first quantitative assessment of the relation between galaxy
morphology and environment was provided by \citet[][D80
hereafter]{dres80}, based on photographic plates of 55 rich galaxy
clusters, obtained using the Las Campanas 2.5m and the Kitt Peak 4m
telescopes. Indeed, it was after this classical paper that the
relation between the local projected density of galaxies in the sky
(computed using the 10 nearest neighbours; $\Sigma_{10}$ hereafter)
and the fraction of galaxies of different morphological types (the
morphology-density relation; $T$-$\Sigma$ hereafter) began to be
considered among the most important empirical evidences of
extragalactic astronomy. In particular, the D80 finding that the
$T$-$\Sigma$ relation holds in both structurally regular (dynamically
``relaxed''?)  and irregular (clumpy, ``non relaxed'') nearby clusters
and its rapid extention to the group and general field environments
\citep{bhav81,desou82,post84}, led the astronomical community to
consider this relation a sort of universal rule in the extragalactic
astronomy. Indeed, the fact that early-type galaxies (Es and S0s; ETGs
hereafter) preferentially dominate dense environments, while late-type
galaxies (Spirals and Irregulars; LTGs hereafter) are more common in
low-density regions has progressively become an out-and-out paradigm
with which to reckon when dealing with problems related to the
evolution of galaxies, from both the phenomenological and theoretical
side. For instance, \citet{evr90} explored a model in which galaxy
morphology is intrinsically related to the peak height of the initial
fluctuations in a CDM universe, finding good agreement with D80 for
the $T$-$\Sigma$ relation of Es.

Besides the existence of the $T$-$\Sigma$ relation in galaxy clusters,
in D80 it is also claimed that this relation is much stronger than the
relation between morphology and cluster-centric distance
(morphology-radius relation; $T$-$R$ hereafter). However, a different
viewpoint was proposed by \citet{WG91}, who re-examined the D80 data,
finding that the strengths of the $T$-$R$ and $T$-$\Sigma$ relations
are comparable. They also suggested that the physical mechanisms
driving the gradients for Es are different from those operating for
spirals and S0s. \citet{WG93} examined again the D80 sample concluding
that the $T$-$R$, normalized to a characteristic radius (R$_c$), is
the fundamental relation driving the morphological fractions of all
types, from spiral galaxies (slightly decreasing toward the center,
then going close to zero near the center), to S0s (moderately
increasing toward the center for R/R$_c<$0.2, then sharply falling
down) and Es (from 10\% in the outer regions to $\sim$16\% at
R/R$_c\sim$1, then rapidly increasing up to 60-70\% in the inner
regions). Moreover, \citet{W95} analysed compact and loose groups and
the field, finding that the $T$-$\Sigma$ is not present in these
environments, just strongly rising near the center of clusters.

The Whitmore \& Gilmore's viewpoint about the prevalence of the
$T$-$R$ over the $T$-$\Sigma$ relation in determining the
morphological fractions in clusters was almost bailed out after the
$T$-$\Sigma$ relation was proved by the MORPHS group \citep[][D97
hereafter]{dres97} to hold also for intermediate redshift clusters
\citep[$z\sim$0.5; see also][in the redshift range
0.1--0.25]{fasa00}. In this redshift range, however, the relation is
found very weak or even absent for irregular clusters. Moreover, we
mention that \citet{treu03} claimed the cluster-centric distance to be
the driving parameter of the ETG fraction in the inner part of the
cluster Cl~0024$+$16 at $z\sim$0.4.

Although structure and morphology are intrinsically different
properties, thus making dangerous to compare visual $T$-$\Sigma$
(morphology) at low $z$ with structural $T$-$\Sigma$ (Sersic index) at
high $z$ \citep{vand08}, in the subsequent decade several authors
studied the $T$-$\Sigma$ relation in different ranges of redshift and
stellar mass. Most of them agreed that: $(i)$ the $T$-$\Sigma$ is
already in place, although in a less strong fashion \citep{post05}, at
$z\sim$0.5, both in groups and (possibly from $z\sim$1) in clusters
\citep{treu03,smit05,desa07,wilm09}; $(ii)$ the evolution of
morphological fractions strongly depends on galaxy mass
\citep{nuij05,oesch10,vulc11}; $(iii)$ in dense environments the ETGs
dominate at all redshifts and their fraction increases over time
\citep{nuij05,smit05,capa07}, mostly because of the growing fraction
of S0s \citep{post05,vulc11}; $(iv)$ the evolution of the $T$-$\Sigma$
relation turns out to be less strong for mass limited than for
magnitude limited samples \citep{hold07,vand07,tasc09}.

Besides the $T$-$\Sigma$ dependence on redshift, some papers have also
analysed the morphological fractions as a function of the global
cluster properties. In particular, \citet{desa07} found a mild
dependence of the morphological fractions on the cluster velocity
dispersion, while the growth of the Spiral/S0 fraction with redshift
was found by \citet{pogg09} to be stronger for low mass than for high
mass clusters. A specific effect of the cluster environment on the
Spiral/S0 fraction is claimed by \citet{calv12}, based on the sharp
enhancement/dearth of S0s/late-types found in clusters at decreasing
redshift, compared to other environments (single galaxies, binary
systems, poor/rich groups).

A totally different and innovative approach to the $T$-$\Sigma$ has been
proposed with the {\it kinematic} morphology-density relation by
\citet[][Paper VII of the Atlas3D collaboration]{capp11}. Rather than
rely on the visual morphology, they classify galaxies according to
their stellar kinematics, dividing ETGs into slow- and fast-rotators
(instead of Es and S0s). In this way and adopting a very local density
estimator, computed just using the 3 nearest neighbours (the Atlas3D
galaxies mainly inhabit low-density environments), they found a $T$-$\Sigma$
cleaner than using classic morphology. They interpret this improvement
as due to the fact that many fast-rotators, visually classified Es are
actually face-on, misclassified S0s. 

WINGS is a multiwavelength photometric and spectroscopic survey of 77
galaxy clusters at 0.04$<$z$<$0.07 \citep[][F06
hereafter]{fasa06}. Clusters were selected in the X-ray from the ROSAT
Brightest Cluster sample and its extension \citep{ebel98,ebel00} and
the X-ray Brightest Abell-type Cluster sample \citep{ebel96}. WINGS
has obtained wide-field optical photometry (BV) for all 77 fields
\citep[][WINGS-OPT]{vare09}, as well as infrared (JK) photometry
\citep[][WINGS-NIR]{vale09}, multi-fiber spectroscopy
\citep[][WINGS-SPE]{cava09} and U-band photometry
\citep[][WINGS-U]{omiz14} for a subset of the WINGS clusters. Surface
photometry and morphology of WINGS galaxies have been obtained through
the purposely devised, automatic tools GASPHOT \citep{dono14} and
MORPHOT \citep[][F12 hereafter]{fasa12}, respectively. A full
description of the WINGS database is given in \citet{more14}.

Exploiting the large and coherent database provided by the WINGS
project, we revisit here the $T$-$\Sigma$ and $T$-$R$ relations in
galaxy clusters of the local Universe. In particular, thanks to the
large galaxy sample, we are in the position of testing how these
relations behave in different ranges of cluster-centric distance (for
$T$-$\Sigma$) and local density (for $T$-$R$), as well as for galaxies
in different stellar mass ranges and for different global cluster
properties.  In Section~\ref{Sec2} we describe the cluster and galaxy
samples. In Section~\ref{Sec3} we give detailed information about the
data used to perform the analysis.  The procedure used to account for
and remove field galaxies is described in
Section~\ref{Sec4}. Section~\ref{Sec5} and Appendix~\ref{App1} are
devoted to characterize the issue related to the limited cluster area
coverage of the WINGS imaging and to investigate the effect of such an
issue on the results we obtain for the $T$-$\Sigma$ and $T$-$R$
relations. These results are presented in Sections~\ref{Sec6}. In
particular, in Sections~\ref{Sec6c} and \ref{Sec6d} we investigate how
$T$-$\Sigma$ and $T$-$R$ depend on the global cluster properties and
on the galaxy stellar mass, respectively. In Section~\ref{Sec7} we
summarize our results and try to outline possible scenarios able to
interpret them.

Throughout the paper we adopt a flat geometry of the Universe with the
following cosmology: $H_0$=70~kms$^{-1}$Mpc$^{-1}$, $\Omega_M$=0.3 and
$\Omega_\Lambda$=0.7.

\section{The sample}\label{Sec2}

\subsection{The cluster sample}\label{Sec2a}

The galaxies used for the present analysis are located in the fields
of 75 clusters belonging to the WINGS-OPT survey (see Table~5 in
F06). The cluster A3164 has been excluded, due to the poor PSF quality
which prevented us from obtaining a reliable morphological
classification. Moreover, when performing the analysis of the $T$-$R$
relation, we excluded the galaxies in the field of clusters A311,
A2665 and ZwCl1261. In fact, for these clusters we do not have a
redshift coverage sufficient to allow a reliable estimation of the
velocity dispersion, which is needed to compute the characteristic
cluster radius R$_{200}$ (as defined in Sec.~\ref{Sec3c}) and
normalize the distance of galaxies from the cluster center.

\subsection{The galaxy sample}\label{Sec2b}

Our starting sample of galaxies roughly concides with that
illustrated in F12. More precisely, the MORPHOT catalog of
39,923 WINGS galaxies has been cross-matched with the catalog of
structural parameters obtained running GASPHOT onto the V-band WINGS
imaging \citep{dono14}. This catalog provides apparent total
magnitudes and structural parameters of galaxies. The apparent
magnitudes have been corrected for galactic extinction using the maps
in \citet{sch98}. The galaxies for which the Sersic index derived by
GASPHOT turned out to coincide with the boundary values of the allowed
search interval (0.5--8) have been excluded from the database. In
fact, in these cases we assume that a failure of the GASPHOT fitting
occurred, thus preventing us from relying on the output GASPHOT
parameters. For the resulting sample of 39,304 objects, putting together the data
from WINGS-SPE \citep{cava09} and from the literature, we gathered
spectroscopic redshifts of 9,361 galaxies, of which 6,297 are cluster
members. The absolute
magnitudes of all galaxies in our sample, but the spectroscopic
non-members, are computed using the average redshifts of clusters
given in \citet{cava09} and the cosmology given in
Section~\ref{secintro}. They are K--corrected using the tables in
\citet{pogg97} and according to the morphological types provided by
MORPHOT (see Section~\ref{Sec3a}).

When comparing our results with those reported in the literature, we
will mainly advert to D80 and D97, which still remain the most
important and cited references on this subject. After accounting for
the different waveband and cosmology, the absolute V-band magnitude
limits (M$_V^{lim}$) given in these papers for the galaxy samples are
$-$19.67 and $-$19.17, respectively. We adopt M$_V^{lim}=-$19.5, which
is also close to the value used for the ENACS sample by
\citet[][;M$_V^{lim}=-$19.57]{thom06}. In this way, we are left with
the final 'reference sample' of 5,504 galaxies, of which 82\% have
measured redshift and 67\% are cluster members.

Throughout the paper, alternative sample partitions and selection
criteria will be tested, so that the actual sample size might vary
from time to time, also depending on the expected field counts
relative to the particular selection. For this reason, we usually
report each time in the following tables the proper sample sizes
before the field removal. Finally, the same above mentioned limit of
luminosity (M$_V^{lim}=-$19.5) is adopted for neighbour galaxies in
the computation of the local densities (Section~\ref{Sec3b}).

\section{The data}\label{Sec3}

All quantities we describe in the followings sub-secions and we use in
the following analysis can be easily retrieved, through the Virtual
Observatory tools, following the recommendations given in
\citet{more14}.

\subsection{Morphology}\label{Sec3a}

The morphological types of galaxies are taken from the database of
WINGS galaxies at:

\noindent
{\tt{http://cdsarc.u-strasbg.fr/viz-bin/VizieR?-meta.foot
    \&-source=J/MNRAS/420/926}}.  This database has been obtained
running MORPHOT (F12) on the V-band WINGS-OPT imaging of galaxies with
isophotal area larger than 200 pixels at the threshold of
2.5$\sigma_{bkg}$ (where $\sigma_{bkg}$ is the standard deviation of
the background). It is worth noticing that none of the WINGS-OPT galaxies with
M$_V\le -$19.5 has been removed from the final sample because of the 200 pixels
cutoff adopted by MORPHOT. This means that our absolute magnitude limited
sample is safe from surface brightness incompleteness (very compact galaxies).

MORPHOT is an automatic tool purposely devised to obtain morphological
type estimates of galaxies in the WINGS survey. It combines a large
set of diagnostics of morphology, easily computable from the digital
cutouts of galaxies. MORPHOT produces two independent estimates of the
morphological type based on: (i) a semi-analytical Maximum Likelihood
technique; (ii) a Neural Network machine. The final estimator has
proven to be almost as effective as the visual classification. In
particular, it has been shown to be able to distinguish between
ellipticals and S0 galaxies with unprecedented accuracy (see Fig.~11
in F12). 

The MORPHOT estimator (MType in the database) is almost coincident
with the Revised Hubble Type defined by \citet[][see Table~1 in
F12]{deva91}. For the purposes in the present paper it is convenient
to split the galaxy sample in three broad morphological types: (i)
ellipticals (E); (ii) lenticulars (S0); (iii) spirals (Sp). With this
convention, the (few) irregular galaxies in our sample are included in
the Sp class and the cD galaxies are included in the E class.

Using a sample of 176 common galaxies belonging to 18 common clusters,
we have compared the broad morphological type (Es/S0s/Spirals)
provided by MORPHOT with the corresponding classifications given in
D80. The agreement turned out to be quite good for spirals
(discrepancy of 2\%), while for Es and S0s we found a discrepancy of
$\sim$20\%, mainly in the sense of an excess of elliptical galaxies in
the MORPHOT classifications with respect to the D80 ones. The most
obvious explanation of such discrepancy is that it originates from the
different material used to inspect clusters: photographic plates in
the original paper of D80 and CCD imaging in the WINGS survey. For
this reason, one of us (AD, indeed!)  has visually (and blindly)
re-classified 25 discordant objects using the WINGS observing material
(CCD). With the new AD classifications the discrepancy turned out to
be reduced down to a physiological amount ($\sim$6\%).
 
\subsection{Local Density}\label{Sec3b}

The projected local galaxy density relative to a given galaxy is
commonly defined as the number of neighbours ($N_n$)of the galaxy per square
megaparsec. According to the prescriptions given in D80, we compute
the projected local densities using the 10 nearest neighbours of the
galaxy with M$_V\le -$19.5. At variance with D80, we use a circle
(rather than a rectangle) to compute the area including the 10 nearest
neighbours. That is: $\Sigma_{10}=10/A_{10}$, where
$A_{10}=\pi R^2_{10}(Mpc)$ and $R_{10}(Mpc)$ is the radius (in Mpc) of
the smallest circle centered on the galaxy and including the 10
nearest neighbours. Numerical simulations have shown that using the
circular area produces a negligible ($\sim$0.03) downward shift of
Log$\Sigma_{10}$ with respect to the values computed using the
rectangular area.

In order to perform the correction for field contamination, rather
than our incomplete membership information, we prefer to use a
statistical correction based on the field counts provided by
\citet{berta} using the deep ESO-Spitzer wide-area Imaging Survey
(ESIS). In particular, we compute for each cluster the number of field
galaxies per $Mpc^2$ ($N_F$) up to the apparent magnitude
corresponding to M$_V^{lim}=-$19.5 at the cluster redshift. Moreover,
in counting the neighbours within a circle centered on a given galaxy,
we statistically account for the objects possibly lost because of the
proximity of the galaxy to the edges of the fied of view (including
the gaps between the chips). We do this dividing the counts by the
ratio $f_c$ ($\le$1) between the area actually covered by the
detector and the circular area.

Starting from the nearest neighbour (i=1) and
progressively including the next nearest one (i=i+1), we compute each
time the circular area $A_i$ (in $Mpc^2$) relative to the neighbour
distance and the corresponding number of neighbours: $N_{i,n}=i
/f_c-N_F\times A_i$. When $N_{i,n}$ overcomes 10, the final
area $A_{10}$ is computed interpolating between $A_i$ and $A_{i-1}$.

\begin{figure}
\vspace{-1truecm}
\resizebox{\hsize}{!}{\includegraphics{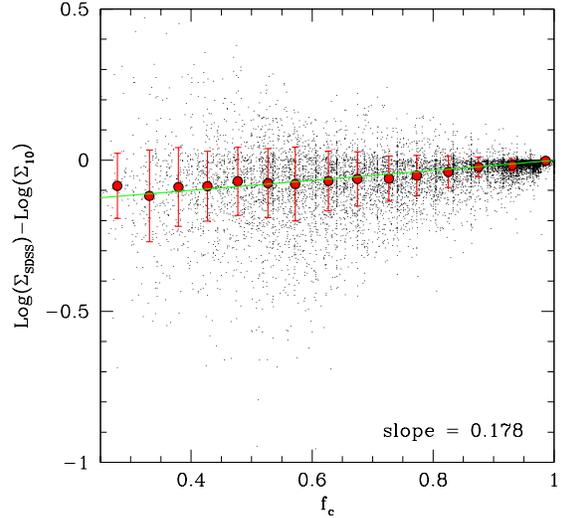}}
\vspace{-2.5truecm}
\caption{Difference between the $\Sigma_{10}$ values obtained with and without 
SDSS information as a function of the coverage fraction.}
\label{fig1}
\end{figure}

For 24 clusters in common with SDSS-DR7, we complement our positional and
photometric database with SDSS information. In particular, we
use the relations in \citet{fuku07} to get the V-band magnitudes of
SDSS galaxies from the $g'$ and $r'$ ones. For this galaxy sample we
are allowed to obtain complete coverages within $A_{10}$ even for
galaxies close to the frame borders, thus being able to compute the $\Sigma_{10}$
values ($\Sigma_{SDSS}$) without correcting for the coverage fraction
($f_c$=1). Figure~\ref{fig1} illustrates how the difference
between the $\Sigma_{10}$ values obtained with and without SDSS information
depend on the coverage fraction. It is evident from Fig.~\ref{fig1}
that for small values of $f_c$ the statistical correction tends to
produce an overestimation of $\Sigma_{10}$. This is likely due to the negative
outward density gradient in clusters, which makes the real number of
galaxies lost because of the proximity to the frame edges, lower than
the number obtained correcting with $f_c$, {\it i.e.}  assuming a
uniform density inside the circles. For this reason the local
densities of galaxies for which no SDSS information is available, have
been statistically corrected using the best-fit relation illustrated
in Fig.~\ref{fig1}: Log$\Sigma_{10}^{corr}$=Log$\Sigma_{10}$+0.178~($f_c$-1).

\subsection{Cluster-Centric Distance}\label{Sec3c}

\begin{figure*}
\includegraphics[width=120mm]{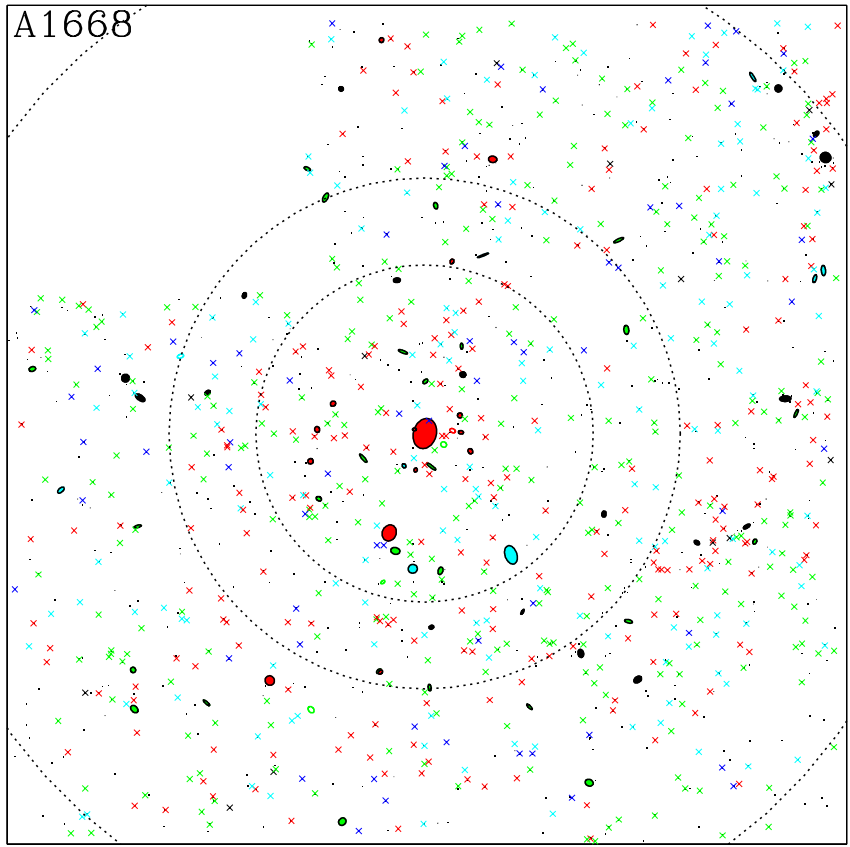}
\caption{Map of the cluster Abell~1668, obtained using the photometric
  data from the WINGS-OPT catalog and the morphologies provided by
  MORPHOT. The different broad morphological classifications are
  reported with different colors: red for Es, green fro S0s, cyan for
  early spirals, blue for late spirals and irregular galaxies. Small
  crosses refer to galaxies fainter than the limit we adopt in our
  analysis (M$_V>-$19.5, adopting the cluster redshift
  distance). Galaxies brighter than this limit are represented by
  ellipses, whose shapes obey the structural parameters provided by
  SExtractor (isophotal area, axial ratio and position
  angle). Spectroscopically confirmed cluster members and galaxies
  without spectroscopic information are indicated by full and empty
  ellipses, respectively, while black symbols (both ellipses and
  crosses) identify galaxies which, according to \citet{cava09}, are
  considered spectroscopic non members. Finally, dashed circles
  correspond to different distances from the BCG: 0.33, 0.5 and 1
  (in R$_{200}$ units).}
\label{figmap}
\end{figure*}
 
We derive two different cluster-centric distances ($R$), depending on
the choice of the cluster center. The first one ($R_B$) assumes the
brightest cluster galaxy (BCG) as cluster center, while the second one
assumes the center of the cluster to coincide with the maximum
intensity of the X-ray emission. In any case, the cluster-centric
distances have to be normalized to some characteristic cluster
size. One class of size estimates is based on top-hat filtered
spherical overdensities. In this model, clusters are expected to be
virialized within regions where the enclosed mean mass density exceeds
the critical density by a factor 200 \citep[][p. 25 and 15,
respectively]{peeb93,peac99}. We assume the radius at which this
overdensity is reached, R$_{200}$, as the characteristic radius of the
cluster. The R$_{200}$ values for our clusters are computed from the
WINGS cluster velocity dispersions ($\sigma$) given in \citet{cava09},
using the formula \citep{finn05}:

$$ {\rm R_{200}=1.73\times{\sigma\over{1000\ km\ s^{-1}}}}\
{1\over{\sqrt{\Omega_\Lambda+\Omega_0(1+{\it z})^3}}}\ {\it h}^{-1}\
\ {\rm (Mpc)} $$.

Therefore, in each cluster, the cluster-centric distances of galaxies ($R$),
to be used in the following analyses, have been obtained dividing
the distance in Mpc by the proper value of R$_{200}$.

In Figure~\ref{figmap} we illustrate, as an example, the map of the
cluster Abell~1668, obtained using the photometric data from the
WINGS-OPT catalog and the morphologies provided by MORPHOT (see the
figure caption for details about symbols).  Similar maps of all
WINGS clusters (76) are available in the online material (see the
electronic version of the paper).

\section{Removing the field galaxies}\label{Sec4}

In order to obtain reliable $T$-$\Sigma$ and $T$-$R$ relations in nearby clusters, it
is important to properly remove the contribution of field galaxies
(mostly background) in each broad morphological class (E/S0/Sp). The
percentages of field galaxies in each class have been estimated using
the Padova-Millennium Galaxy and Group Catalogue
\citep[][PM2GC]{calv11,calv12}, consisting of a spectroscopically
complete sample of galaxies at 0.03$\leq z\leq$0.11 brighter than
M$_B=-$18.7. This sample is sourced from the Millennium Galaxy
Catalogue \citep{lisk03,driv05}, a B-band contiguous equatorial
survey, complete down to B=20 and representative of the general field
population in the local Universe. The advantage of using PM2GC data
is that both the instrumental set up of the imaging (WFC@INT) and the
tool used to estimate the morphological types (MORPHOT) are the same
for both PM2GC and WINGS galaxies. This guarantees a full
morphological consistency between the two samples.

\begin{figure}
\vspace{-1truecm}
\resizebox{\hsize}{!}{\includegraphics{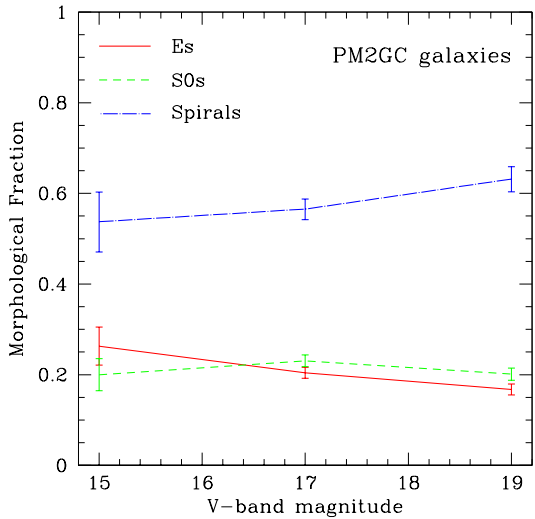}}
\vspace{-2.5truecm}
\caption{Morphological fractions of PM2GC galaxies as a function of
  the apparent V-band magnitude. Error bars report the poissonian uncertainties}
\label{fig2}
\end{figure}

\begin{figure}
\vspace{-1truecm}
\resizebox{\hsize}{!}{\includegraphics{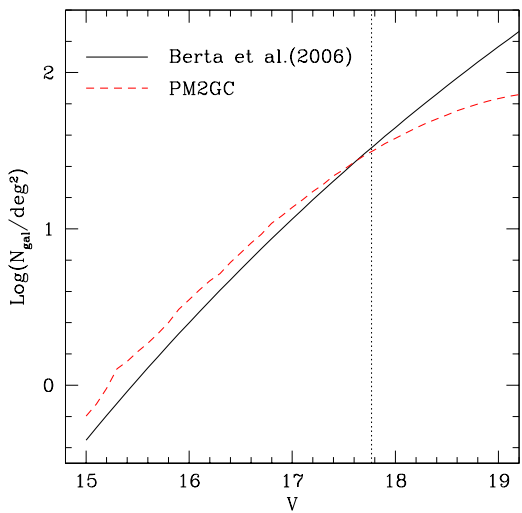}}
\vspace{-2.5truecm}
\caption{Counts of field galaxies from \citet{berta} 
  and from PM2GC. The vertical dotted line corresponds to the apparent
magnitude at which we have computed the morphological fractions (V=17.77).}
\label{fig3}
\end{figure}

Figure~\ref{fig2} reports the morphological fractions of the PM2GC
galaxy sample in different bins of apparent magnitude. Even if the
expected increase of the E fraction and decrease of the Sp fractions
are clearly detectable, Fig.~\ref{fig2} shows that the percentages
E/S0/Sp depends weakly on the apparent magnitude. In our analysis we
assume E/S0/Sp = 19.3($\pm$3.1)\%/21.7($\pm$2.7)\%/59($\pm$5.3)\%.
These values correspond to the morphological fractions of PM2GC
galaxies with V$\le$17.77, this limit including 95\% of the WINGS
galaxies in our reference sample (see below).

In Figure~\ref{fig3} the global field counts in the V-band from the
PM2GC galaxy sample are compared with the counts given by
\citet{berta} in the same waveband and in the magnitude range of
interest for us. The slight systematic excess of the PM2GC counts for
bright galaxies with respect to the counts by \citet{berta} is likely
due to the presence in the PM2GC field of many galaxy groups, which,
because of the criteria adopted in the choice of the field, are
probably lacking in the ESIS database used by \citet{berta}. On the
other hand, the sharp decline of the PM2GC counts towards faint
galaxies clearly reveals the incompleteness of the PM2GC galaxy sample
for which a morphological classification has been possible through
MORPHOT (isophotal area larger than 200 pixels at the threshold of
2.5$\sigma_{bkg}$). Nevertheless, it is noticeable that the two counts
curves assume roughly the same value (1.51, that is $\sim$32 galaxies
per square degree) at V$\le$17.77, which is the value chosen above to
compute our morphological fractions. All things considered, the
statistical counts of field galaxies suggest us to remove in our
analyses 32 galaxies per square degree, in particular: 6~Es (19.3\%),
7~S0s (21.7\%) and 19~Spirals (59\%).

According to Section~\ref{Sec2b}, in our galaxy sample the global
spectroscopic coverage and cluster membership, up to the adopted limit
of absolute V-band magnitude ($-$19.5), are 82\% and 67\%,
respectively. These rather high percentages allowed us to deal with
the problem of field galaxies using a strategy complementing the
statistical field counts with the membership information. In
particular, we first removed from the sample those galaxies which,
according to the redshift information and the membership criteria
described in \cite{cava09}, are not cluster members. Then, if the
number of galaxies removed from each broad morphological class
(Es/S0s/Sp) turned out to be lower than the expected field counts
relative to that class, we further subtracted from the residual the
quantity needed to reach the field counts expected from \citet{berta}.
In this way, it might happen that the actual number of removed
galaxies exceeds the number expected from the field counts. In any
case, we have verified that the differences between the $T$-$\Sigma$
relations obtained using our mixed strategy and a `pure' statistical
approach are practically negligible.

\section{The cluster area coverage issue}\label{Sec5}

Having at our disposal a large galaxy sample in a well defined,
complete cluster sample, our strength lies in the opportunity to take
advantage of a good statistics. For this reason, in the present
analysis we mostly use galaxies in the whole cluster sample. At times,
however, in order to perform some particular analysis, we use galaxies
belonging to a sub-sample of clusters (selected, for instance, on the
basis of X-ray emission, velocity dispersion or subclustering), or
galaxies obeying some constraint (for instance on the stellar mass).

\begin{figure}
\vspace{-1.5truecm}
\resizebox{\hsize}{!}{\includegraphics{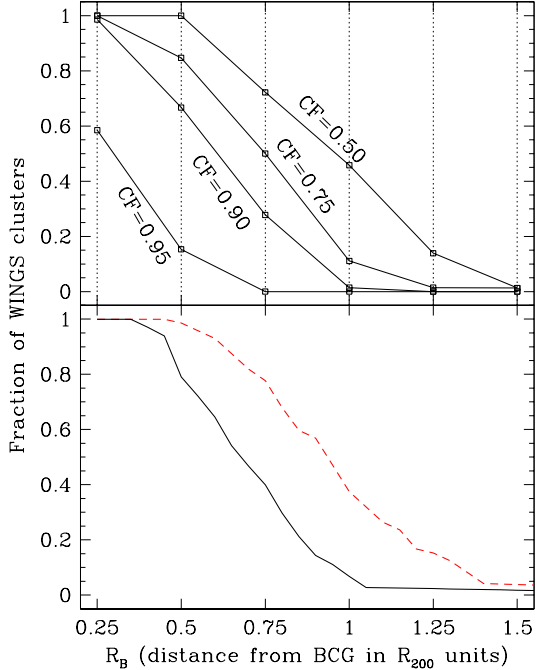}}
\vspace{-1truecm}
\caption{{\bf Upper Panel}: The fraction of WINGS clusters with
  circular coverage fraction greater than some given value (0.50,
  0.75, 0.90 and 0.95) as a function of the radius (in units of
  R/R$_{200}$) of the circular area centered on the BCG. 
  {\bf Bottom Panel}: The solid line reports the fraction of WINGS
  clusters whose equivalent radius (see text) exceeds the value of
  $R_B$ reported in the abscissa.  The dashed line (red in the
  electronic version) illustrates the fraction of WINGS clusters for
  which the largest clustercentric galaxy distance ($R_B$) exceeds the
  value given in the abscissa.}
\label{fig4}
\end{figure}

On the other hand, our weakness lies in the limited and not always
regular cluster area covered by the WINGS images. The upper panel of
Figure~\ref{fig4} shows that, for R$_B$=0.5, 0.75 and 1.0, the WINGS
images cover more than 90\% of the circular (BCG centered) area in
just $\sim$66\%, 28\% and 2\% of the clusters, respectively, while a
coverage fraction (CF hereafter) of 0.75 at the same radii is reached
in $\sim$85\%, 50\% and 11\% of the clusters. Similar CFs are found if
we assume the cluster centers to coincide with the maximum intensity
of the X-ray emission. An alternative way to illustrate the coverage
problem is provided by the bottom panel of Fig.~\ref{fig4}. In this
panel the solid line reports the fraction of WINGS clusters whose
equivalent radius R$_{eq}=\sqrt{A_{200}/\pi}$ ($A_{200}$ being the
area covered by the WINGS images in units of R$^2_{200}$) exceedes the
value of $R_B$ reported in the abscissa, while the dashed line (red in
the electronic version) reports the fraction of WINGS clusters for
which the galaxy with the largest clustercentric distance has a $R_B$
exceeding the value given in the abscissa.

Even if one of the goals of the present paper is the analysis of the
morphological fractions as a function of the clustercentric distrance
($T$-$R$), from previous studies we know that the spiral fraction inside a
given (BGC or X-ray centered) cluster area is an increasing function
of the radius. Thus, the limited cluster area coverage of the WINGS
images in comparison with other analyses \citep[{\it e.g.:}
D80;][]{goto03,thom06}, is expected to result in a reduced global
fraction of spirals.

Besides this obvious effect, additional biases due to the limited
cluster area coverage of our data could affect the $T$-$R$ and
$T$-$\Sigma$ relations we present here.

Concerning the first relation ($T$-$R$), there is no reason to believe
that any galaxy included in our sample, with clustercentric distance
($R$) not completely covered by our imaging, should be drawn from a
morphological distribution different from the average distribution at
that $R$. Therefore, the $T$-$R$ should be unaffected by both the
limited cluster area coverage of our images, and their irregular shape
(for INT imaging, in particular). For the $T$-$R$, the above issues
should just translate into a limited extension of the relation and a
relatively large poissonian uncertainty of its outer part.

Trying to quantify the possible biases affecting the $T$-$\Sigma$
relation because of the limited cluster area coverage of our imaging, in
Appendix~\ref{App1} we present three different tests. The first two
tests investigate the effects on $T$-$\Sigma$ of two different issues:
the irregular shape of WINGS images and their limited coverage of the
cluster area. The third test, performed through quite simple numerical
simulations, is again aimed at investigating the effect of the limited
cluster coverage on the $T$-$\Sigma$ in different ranges of
clustercentric distance. Moreover, since the simulations require some
hypothesis about the driving parameter ($\Sigma$ or $R$) of the
morphological fractions, they could also provide useful indication
with regards to such parameter (see Section~\ref{App1c} in
Appendix~\ref{App1}).

\section{Results: The $T$-$\Sigma$ and $T$-$R$ relations in WINGS clusters}\label{Sec6} 

In Fig.~\ref{fig4} we have shown that, due to the limited field of
view of the cameras used in the WINGS optical survey, the cluster area
coverage of our cluster images turns out to be relatively small. As
expected, this makes the global morphological content of our galaxy
sample quite different from that reported in previous analyses which
made use of much larger field of view imaging. In particular, we find
E/S0/Sp$\sim$33\%/44\%/23\% \citep[see also][and \citealt{vulc11}, with
a slightly different sample selection]{pogg09}, to be compared with
the morphological fractions given by D80 ($\sim$18\%/41\%/41\%) and
\citet[][$\sim$19\%/47\%/34\%]{thom06}. It is worth stressing that,
supported by the comparison between the MORPHOT and D80 morphological
classifications (see Section~\ref{Sec3a}), we entirely attribute to
the limited cluster area coverage of the WINGS imaging the above
discrepancy of global morphological fractions.

\noindent
Let's now indicate with $F_{fE}$/$F_{fS0}$/$F_{fSp}$=0.193/0.217/0.59
the morphological fractions we have previously found for field
galaxies, and with $N_{tot}$ and $N_f$ the total number of galaxies
and of field galaxies relative to the particular selection (of
clusters and/or galaxies and/or range of $R$ or $\Sigma_{10}$) we are
testing in the analysis of the $T$-$\Sigma$ and $T$-$R$ relations. For
each bin ($i$) of Log$\Sigma_{10}$ or $R_B$ and for each broad
morphological type (T=E/S0/Sp), we compute the morphological fractions
as:

\begin{center}
\begin{equation}\label{eq1}
$$F^i_T={N^i_T-F_{fT}N^i_f\over{N^i-N^i_f}}.$$
\end{equation}
\end{center}

In these expressions, the symbols $N$ and $F$ indicate galaxy numbers
and fractions, respectively. In particular, $N^i$ and $N^i_T$ are the
total number of galaxies and the number of galaxies with broad
morphological type T in the $i^{th}$ bin, while we indicate with
$N^i_f$ the product $N^i\times F_f$ and with $F_f$ the fraction
$N_f/N_{tot}$. In the above formulation of $F^i_T$ we assume
that the number of field galaxies per square degree and the global
morphological field ratios $F_{fT}$ do not depend on the local
density. In the following figures we just plot the bins where the
number of cluster galaxies ($N^i-N^i_f$) is greater than 10.

\subsection{$T$-$\Sigma$ relation}\label{Sec6a}

\begin{figure}
\vspace{-0.5truecm}
\includegraphics[width=91mm]{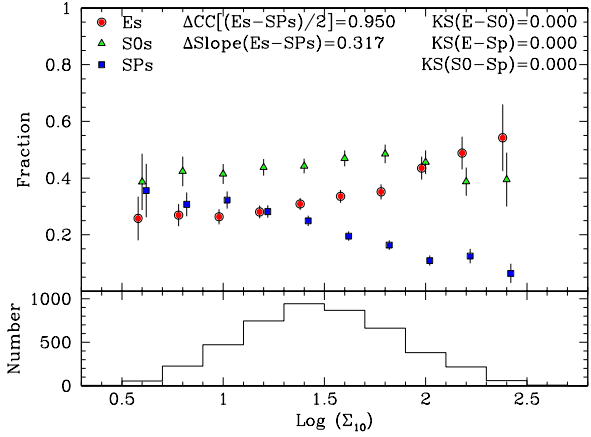}
\vspace{-0.5truecm}
\caption{$T$-$\Sigma$ (upper panel) and histogram of local density (lower
  panel) for the whole 'reference sample' of 5504 WINGS galaxies with
  M$_V\le -$19.5. Full dots, triangles and squares (red, green and blue
  in the electronic version of the paper) refer to Es, S0s and
  Spirals, respectively. The error bars correspond to poissonian
  uncertainties. Details about the coefficients reported in the upper
  panel (KS, $\Delta$CC and $\Delta$Slope) are given in the text.}
\label{fig5}
\end{figure}

Figure~\ref{fig5} illustrates the $T$-$\Sigma$ relation for the whole
reference sample of 5,504 WINGS galaxies (see also
Table~\ref{tabres}), adopting a bin size of 0.2 in
Log$\Sigma_{10}$. In the upper panel of the figure and in the
following $T$-$\Sigma$ (and $T$-$R$) plots, we report a few
coefficients, through which we try to quantify the differences among
the Fraction-Log$\Sigma_{10}$ (Fraction-$R_B$ for the $T$-$R$)
relations relative to Es, S0s and Spirals. In particular, for each
pair of broad morphological types, we report the two-samples
Kolmogorov-Smirnov probability (2S-KS hereafter) that the two
distributions of Log$\Sigma_{10}$ are drawn from the same parent
population. Moreover, we compare the weighted $T$-$\Sigma$ of Es and
Spirals, reporting the coefficients $\Delta$CC and $\Delta$Slope. They
give respectively the halved difference of Correlation Coefficients
and the Slope difference between the $T$-$\Sigma$ of Es and Spirals
assuming a linear fit. These values, together with their expected
r.m.s. uncertainties, are also reported in Table~\ref{tabres}.

Figure~\ref{fig5} clearly shows that for the WINGS cluster galaxies we
recover the classical $T$-$\Sigma$ relation: at increasing local density,
the fractions of Es and Spirals increase and decrease,
respectively. The correlations are very strong in both cases, as
indicated by the coefficients $\Delta$CC and $\Delta$Slope, while the
KS(E-Sp) probability indicates that Es and
Spirals are quite distinct from each other. It is also worth noticing
that the fraction of S0s seems not to depend at all on the local
density, as found also by D97. In addition, the KS analysis suggests
that the S0s constitute a population well distinct from both Es and
Spirals. The first row of Table~\ref{tabres} (columns 2, 3 and 4)
reports the values of the coefficients, together with their r.m.s
uncertainties.

\begin{figure*}
\vspace{-0.5truecm}
\hspace{-0.8truecm}
\includegraphics[width=91mm]{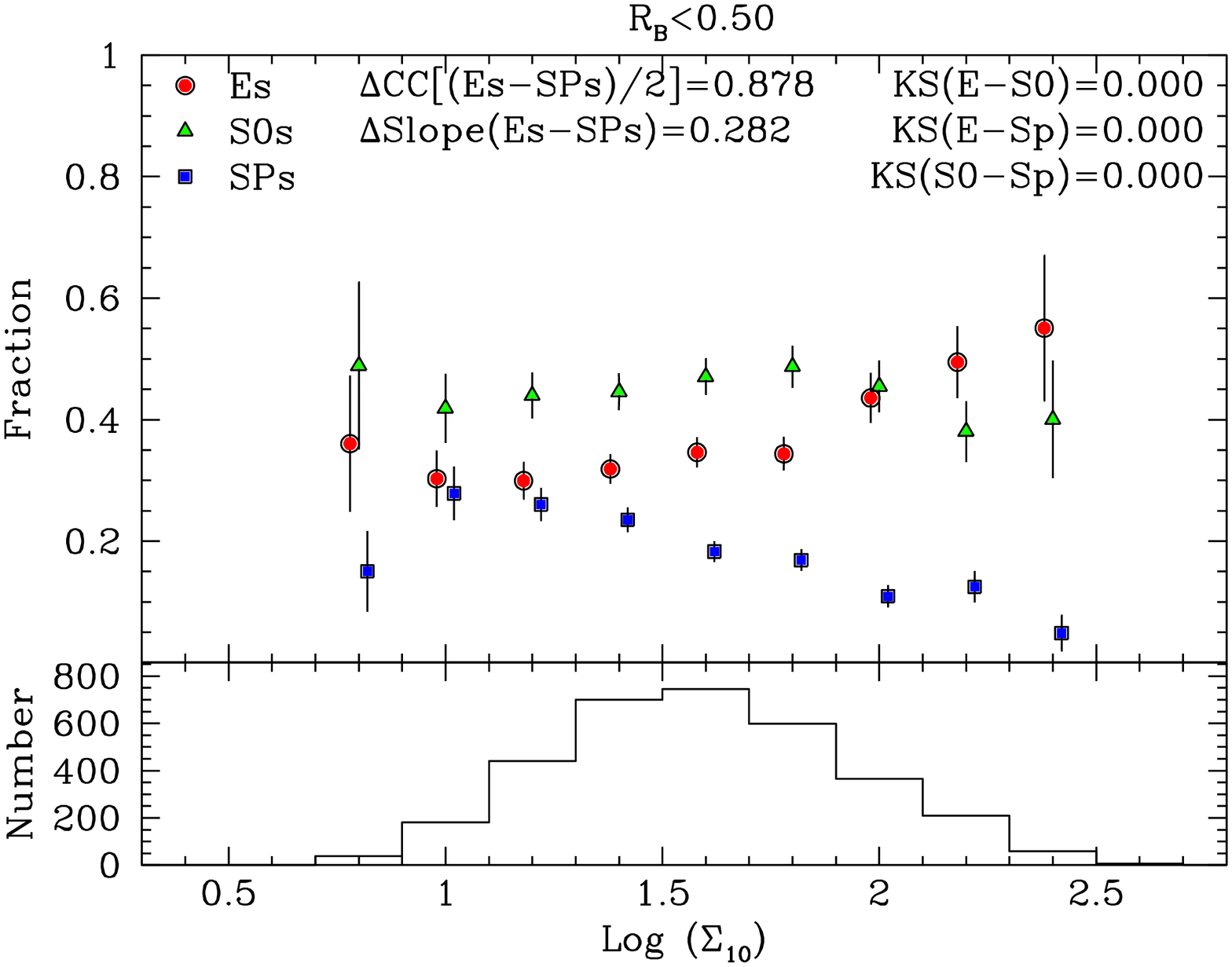}
\includegraphics[width=91mm]{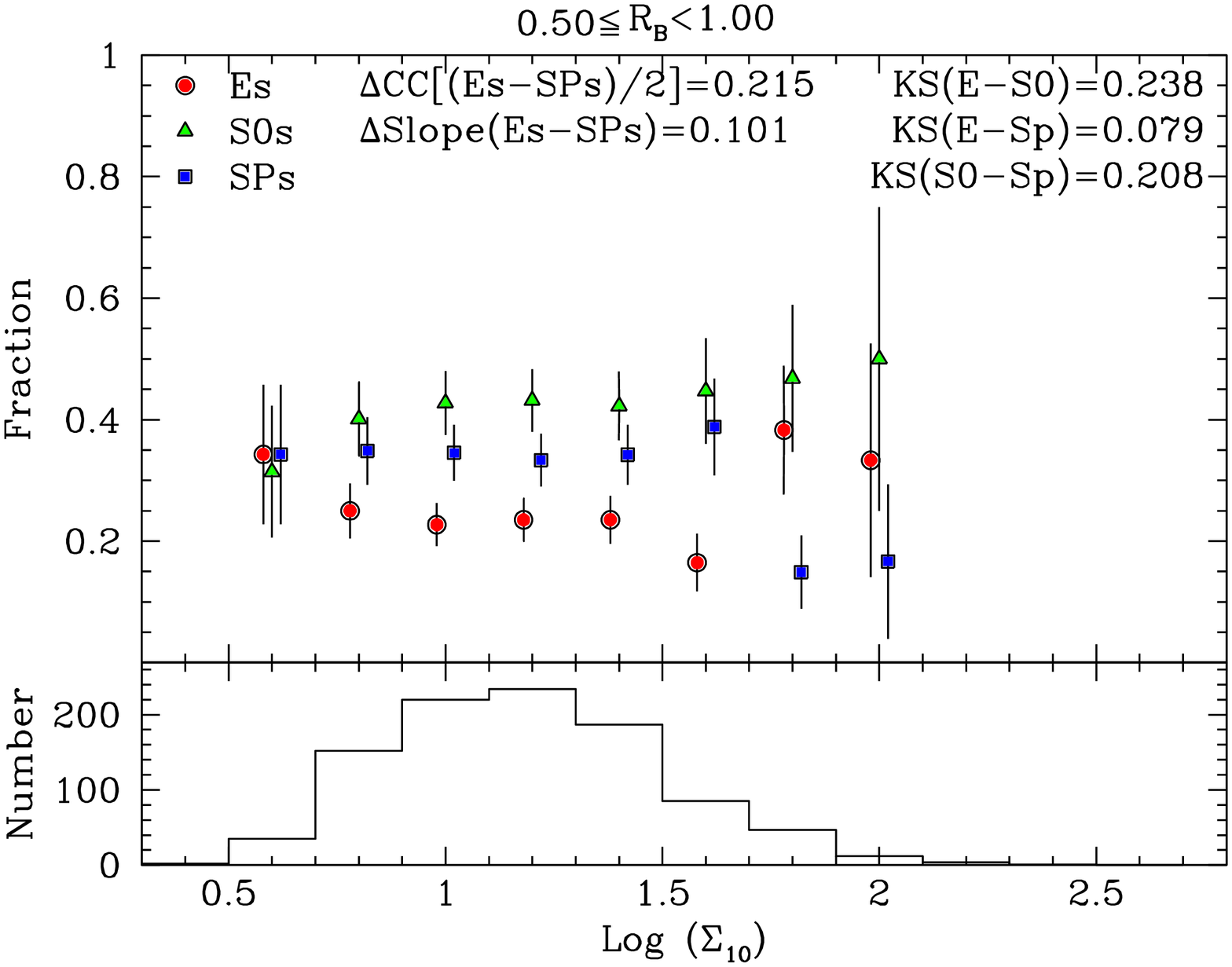}
\caption{$T$-$\Sigma$ for WINGS galaxies with $R_B<$0.5 (in units of
  R$_{200}$, left panel) and 0.5$\leq R_B<$1 (right panel). The meaning
  of the symbols is as in Fig.~\ref{fig5}.}
\label{fig6}
\end{figure*}

Let's now check out whether and how the $T$-$\Sigma$ relation depends
on the position of the galaxy inside the cluster, in particular on the
clustercentric distance relative to the BCG ($R_B$). Figure~\ref{fig6}
illustrates the $T$-$\Sigma$ in two different ranges of $R_B$. The
left panel of the figure refers to the central region of the clusters
($R_B<$0.5), while the right panel refers to the $R_B$ interval
(0.5--1). It stands out that, while in the central part of the cluster
the $T$-$\Sigma$ is quite strong, when moving outside it seems to
become very weak or even absent. This is formally confirmed by the
$\Delta$CC and $\Delta$Slope values, as well as by the 2S-KS
probabilities obtained in the two $R_B$ intervals.  These values are
reported in Table~\ref{tabres} (rows 2 and 3; columns 2, 3 and 4).

Table~\ref{tabres} also reports, in rows 4, 5 and 6 (columns 2, 3 and 4),
the results we obtain splitting the $R_B$ interval (0--1) into three
parts, while in the row 7 we report the data relative to the $R_B$
interval (0.33--1), which we refer to in
Figure~\ref{fig9}. Table~\ref{tabres} clearly shows that the region
where the $T$-$\Sigma$ actually operates is the very inner cluster
region ($R_B<$1/3). The high significance found for the $T$-$\Sigma$
for the whole galaxy sample (see Fig.~\ref{fig5}) is just due to the
fact that the inner regions are the most densely populated in the
clusters.

\begin{table*}
  \caption{Coefficients $\Delta$CC and $\Delta$Slope for $T$-$\Sigma$ and
    $T$-$R$ in different intervals of $R_B$ and Log$\Sigma_{10}$, respectively. The
    table also report the 2S-KS probabilities that the Log$\Sigma_{10}$ or $R_B$
    distributions of elliptical and spiral galaxies are drawn from the
    same parent population for the $T$-$\Sigma$ and $T$-$R$ relations, respectively. The
    number of galaxies used in each case is reported below the $R_B$ or 
    Log$\Sigma_{10}$ intervals, while the expected r.m.s. uncertainties of the 
    coefficients $\Delta$CC and $\Delta$Slope are reported below the 
    relative values.} 
\begin{tabular}{cccccccc}
\hline
\multicolumn{4}{c}{$T$-$\Sigma$} & \multicolumn{4}{c}{$T$-$R$} \\
\hline
$R_B$ &  $\Delta$CC & $\Delta$Slope & KS(E-Sp)\ \ \  &\ \ \ Log$\Sigma_{10}$ & $\Delta$CC & $\Delta$Slope & KS(E-Sp) \\
\hline
 All & 0.950 & 0.317 & 0.000 & 0--3 & 0.942 & 0.619 & 0.000 \\
 $^{(5504)}$&$^{(0.024)}$&$^{(0.025)}$&&$^{(5187)}$&$^{(0.026)}$&$^{(0.053)}$&\\
 0--0.5 & 0.878 & 0.282 & 0.000 & 0--1.45 & 0.869 & 0.478 & 0.000 \\
 $^{(3813)}$&$^{(0.054)}$&$^{(0.041)}$&&$^{(2567)}$&$^{(0.055)}$&$^{(0.070)}$&\\
 0.5--1 & 0.215 & 0.101 & 0.079 & 1.45--3 & 0.878 & 0.602 & 0.000 \\
 $^{(1321)}$&$^{(0.195)}$&$^{(0.083)}$&&$^{(2620)}$&$^{(0.062)}$&$^{(0.105)}$&\\
\hline
0--0.33 & 0.910 & 0.342 & 0.000 & 0--1.2 & 0.762 & 0.375 & 0.001 \\
 $^{(2538)}$&$^{(0.047)}$&$^{(0.048)}$&&$^{(1331)}$&$^{(0.094)}$&$^{(0.080)}$&\\
0.33--0.66 & 0.132 & 0.018 & 0.280 & 1.2--1.8 & 0.953 & 0.578 & 0.000 \\
 $^{(2026)}$&$^{(0.195)}$&$^{(0.042)}$&&$^{(2828)}$&$^{(0.025)}$&$^{(0.059)}$&\\
 0.66--1 & 0.145 & 0.113 & 0.120 & 1.8--3 & 0.709 & 0.522 & 0.000 \\
 $^{(570)}$&$^{(0.228)}$&$^{(0.134)}$&&$^{(1028)}$&$^{(0.171)}$&$^{(0.181)}$&\\
 0.33--1 & 0.332 & 0.075 & 0.089 &&&&\\
 $^{(2596)}$&$^{(0.184)}$&$^{(0.041)}$&&&&&\\
\hline
\end{tabular}
\label{tabres}
\end{table*}

\subsection{$T$-$R$ relation}\label{Sec6b}

In the analysis of the $T$-$R$ relation we assume again the cluster
center to coincide with the position of the BCG ($R\equiv R_B$), while
in this case we adopt a bin size of 0.15 for $R_B$.  We note, however,
that both the previous and the following results remain practically
unchanged if we adopt a different bin sizes and/or assume the cluster
center to coincide with the maximum intensity of the X-ray emission.

\begin{figure}
\vspace{-0.5truecm}
\includegraphics[width=91mm]{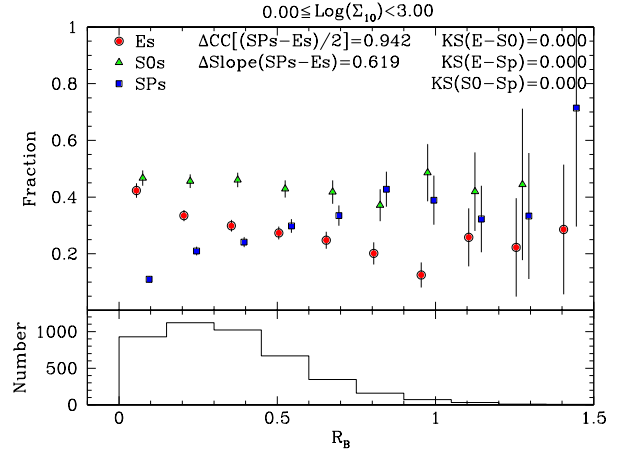}
\vspace{-0.5truecm}
\caption{$T$-$R$ (upper panel) and histogram of $R_B$ (lower panel) for the
  whole sample of 5187 WINGS galaxies with M$_V\le -$19.5 for which the
  value of $R_B$ in units of R$_{200}$ is available. The meaning of the
  symbols is as in Fig.~\ref{fig5}.}
\label{fig7}
\end{figure}

\begin{figure*}
\vspace{-0.5truecm}
\hspace{-0.8truecm}
\includegraphics[width=91mm]{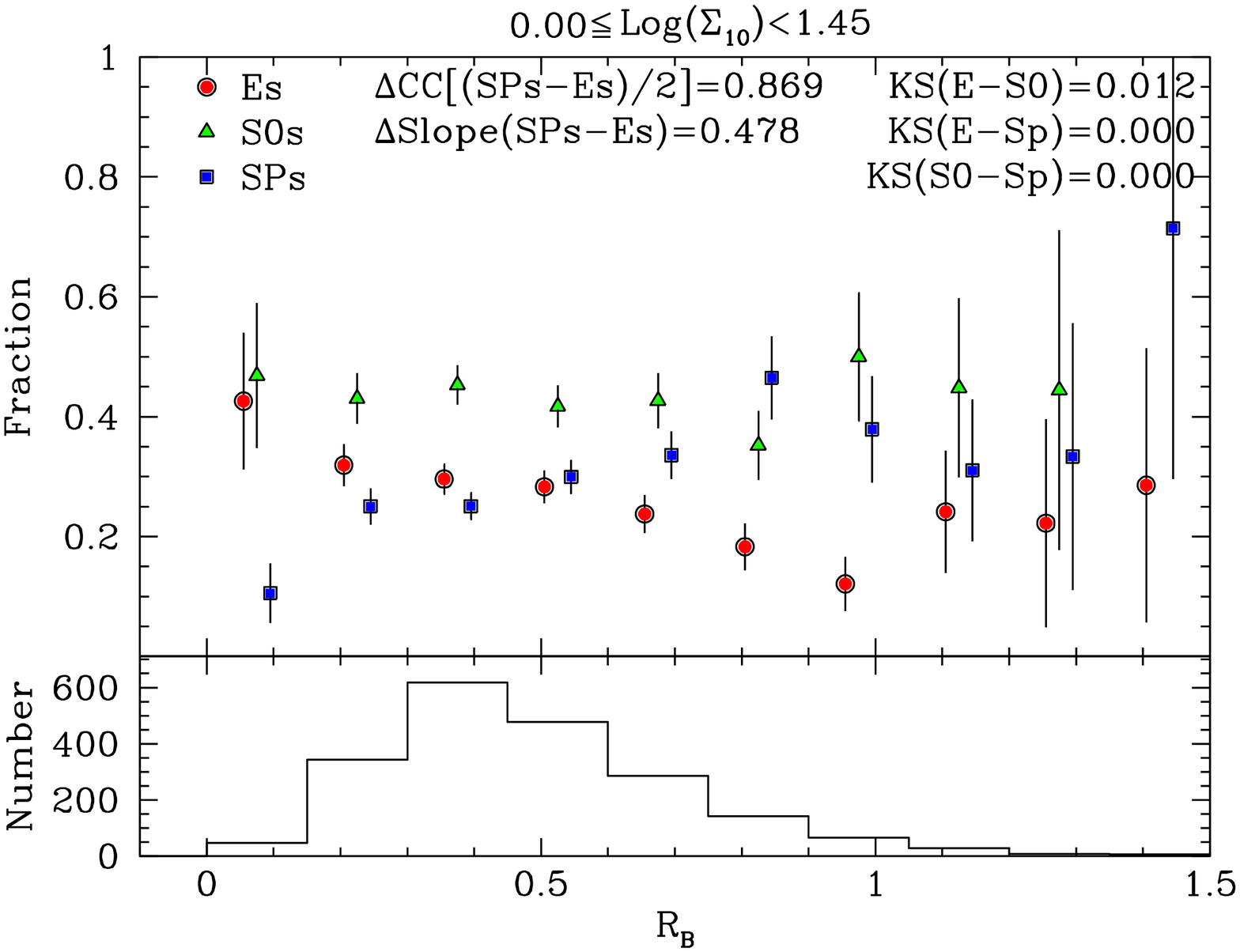}
\includegraphics[width=91mm]{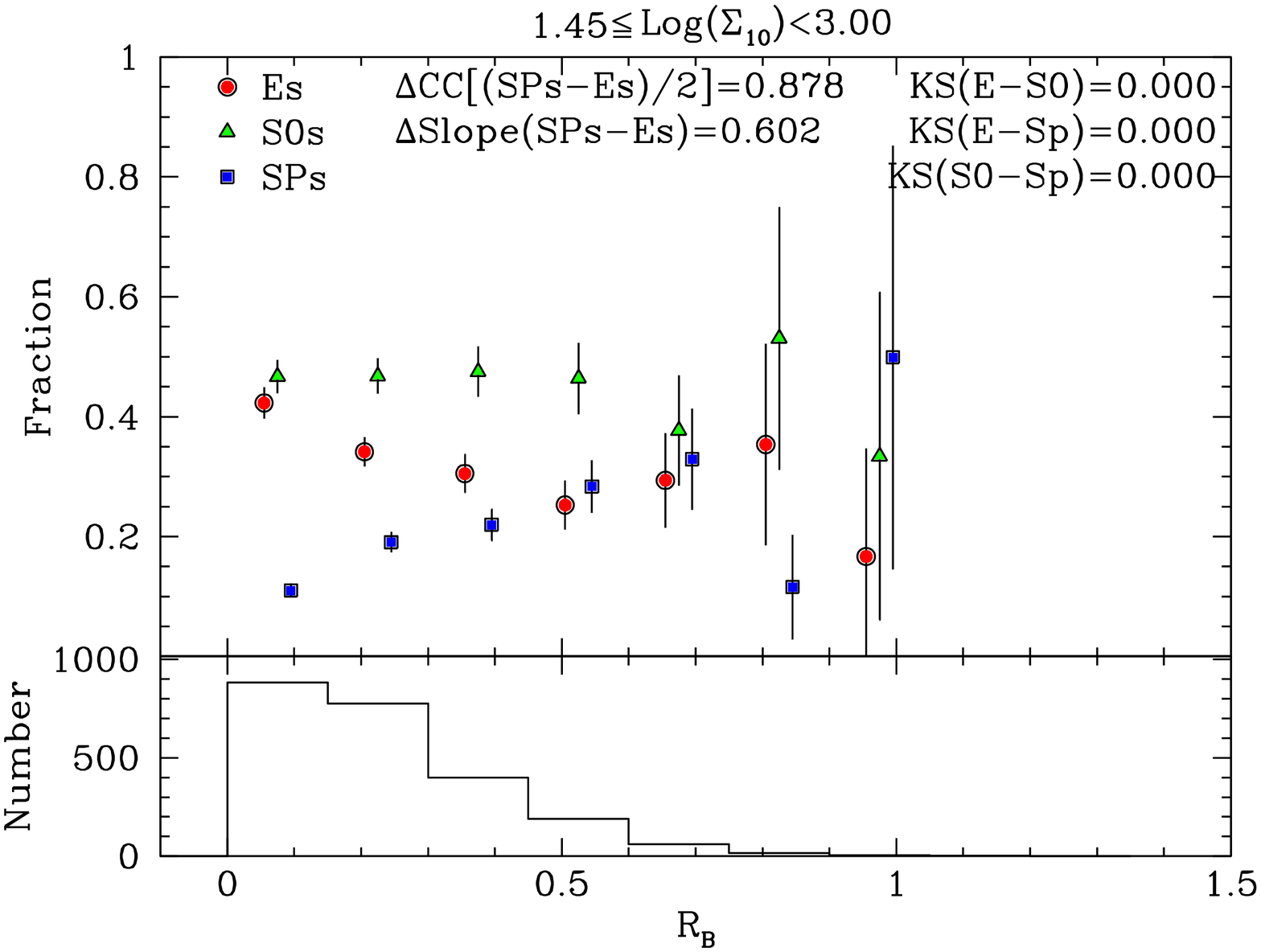}
\caption{$T$-$R$ for WINGS galaxies with Log$\Sigma_{10}<$1.45 (median value of
  the $\Sigma_{10}$ distribution; left panel) and 1.45$\leq$Log$\Sigma_{10}<$3 (right
  panel). The meaning of the symbols is as in Fig.~\ref{fig5}.}
\label{fig8}
\end{figure*}

Figures~\ref{fig7} and \ref{fig8} are similar to Figs.~\ref{fig5} and
\ref{fig6}, respectively, but refer to the $T$-$R$ relation for the global sample
of WINGS galaxies (Fig.~\ref{fig7}) and for galaxies in two different
ranges of the local density (Figs.~\ref{fig8}). Columns 6, 7 and 8 of
Table~\ref{tabres} report the values of the coefficients in these
cases, as well as the values obtained splitting the Log$\Sigma_{10}$
interval (0--3) into three parts.

This table and the two above mentioned figures show that, at variance
with the $T$-$\Sigma$ relation, where the morphological correlations
turn out to be significant just in the inner regions of clusters, the
$T$-$R$ relation turns out to be always significant in the cluster
environment (at least out to $R \sim R_{200}$), irrespectively of the
local density regime. This is highly suggestive that the $T$-$R$
relation is more robust than the $T$-$\Sigma$ one, and that the
clustercentric distance is actually the driving parameter of the
morphological fractions in clusters. The above conclusions seem to
overturn the widespread paradigm of the Morphology-Density relation,
making the old, alternative point of view by \citet[][see also
\citealt{W95}]{WG91,WG93} to come up again. In Section~\ref{Sec7} we
briefly discuss this point, trying to interpret the strong outwards
weakening of $T$-$\Sigma$ as due to some mechanism of morphological
broadening/reshuffling, which could operate in the intermediate/outer
regions of nearby clusters. In the meantime, it is worth mentioning
that the robustness of our results seems to be supported by the tests
and simulations presented in Appendix~\ref{App1}. We are aware that
they cannot fully remedy the lack of data at large clustercentric
distances. Still, waiting for the conclusion of our program for very
wide-field (1$^\circ\times$1$^\circ$) imaging and spectroscopy of a
representative sub-sample of the WINGS clusters\footnote{We are
  presently gathering V-, B- and u'-band OmegaCam@VST imaging and
  AAOmega$+$2dF@AAT spectroscopy for 54 clusters of the WINGS original
  sample.}, we believe the results we present here should be a good
approximation of those we will obtain with a more complete galaxy
sample.

\subsection{$T$-$\Sigma$ and $T$-$R$ for different global cluster properties}\label{Sec6c}

We now seek for possible dependencies of the $T$-$\Sigma$ and
$T$-$R$ relations on global cluster properties. In particular, we
investigate the dependencies on the X-ray luminosity (Log$L_X$) and on
the velocity dispersion of galaxies in the clusters (Log$\sigma$),
both quantities being somehow linked to the total cluster
mass. Moreover, we analyse how the presence of substructures in the
clusters influences the strenght of the $T$-$\Sigma$ and $T$-$R$
relations.

\subsubsection{Cluster mass}\label{Sec6c1}

Cluster velocity dispersions were computed combining WINGS and
literature redshifts. For all but one cluster, they are based on more
than 20 spectroscopic members, with an average of 92 spectroscopic
members per cluster \citep{cava09}.
The X-ray luminosities (0.1 2.4 keV) from \citet{ebel96,ebel98,ebel00}
have been converted to the cosmology used in this paper.

The WINGS clusters cover a wide range of $\sigma$ , typically between
500 and 1100~km~s$^{-1}$, and $L_X$, typically
0.2-5$\times$10$^{44}$~erg~s$^{-1}$.

Tables ~\ref{tabsig} and \ref{tablx} are similar to Table~\ref{tabres}
(first three rows), but they split the WINGS cluster sample in two
different intervals of Log$\sigma$ and Log$L_X$, respectively. The
splitting values are choosen trying to balance (as much as possible)
both the width and the number of galaxies of the intervals themselves.

In spite of the expected (sometimes large) uncertainties of the
coefficients $\Delta$CC and $\Delta$Slope, Tables ~\ref{tabsig} and
\ref{tablx} suggest that the results we found for the global WINGS
cluster sample do not depend either on the velocity dispersion, or on
the X-ray luminosity of clusters. In fact, in both cases, the galaxy
sub-samples obtained splitting the global WINGS cluster sample, again
show the tendency of the $T$-$\Sigma$ relation to hold just in the
inner cluster regions, while the strength of the $T$-$R$ relation does
not seem to depend on the local density regime.

\begin{figure*}
\vspace{-0.5truecm}
\hspace{-0.8truecm}
\includegraphics[width=91mm]{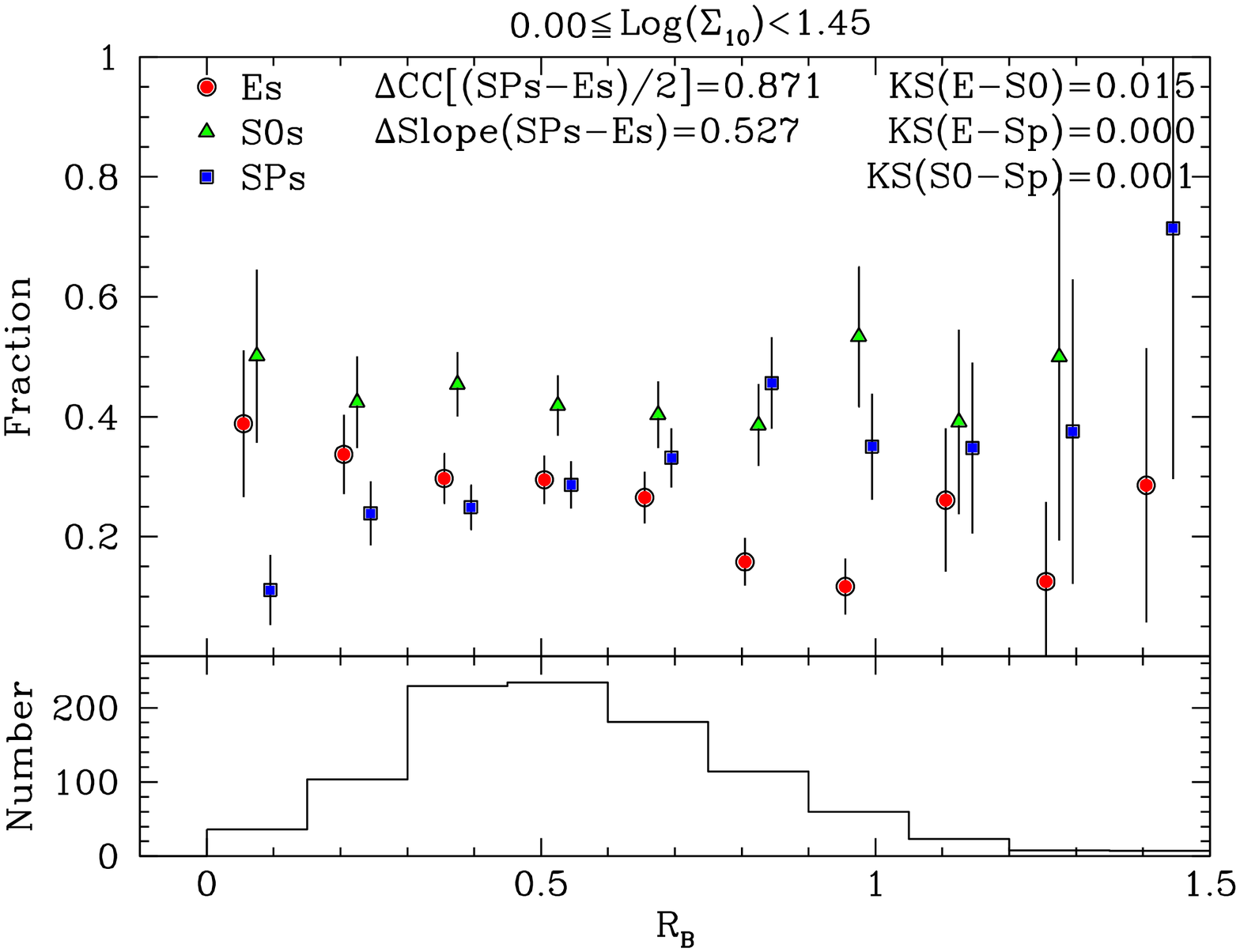}
\includegraphics[width=91mm]{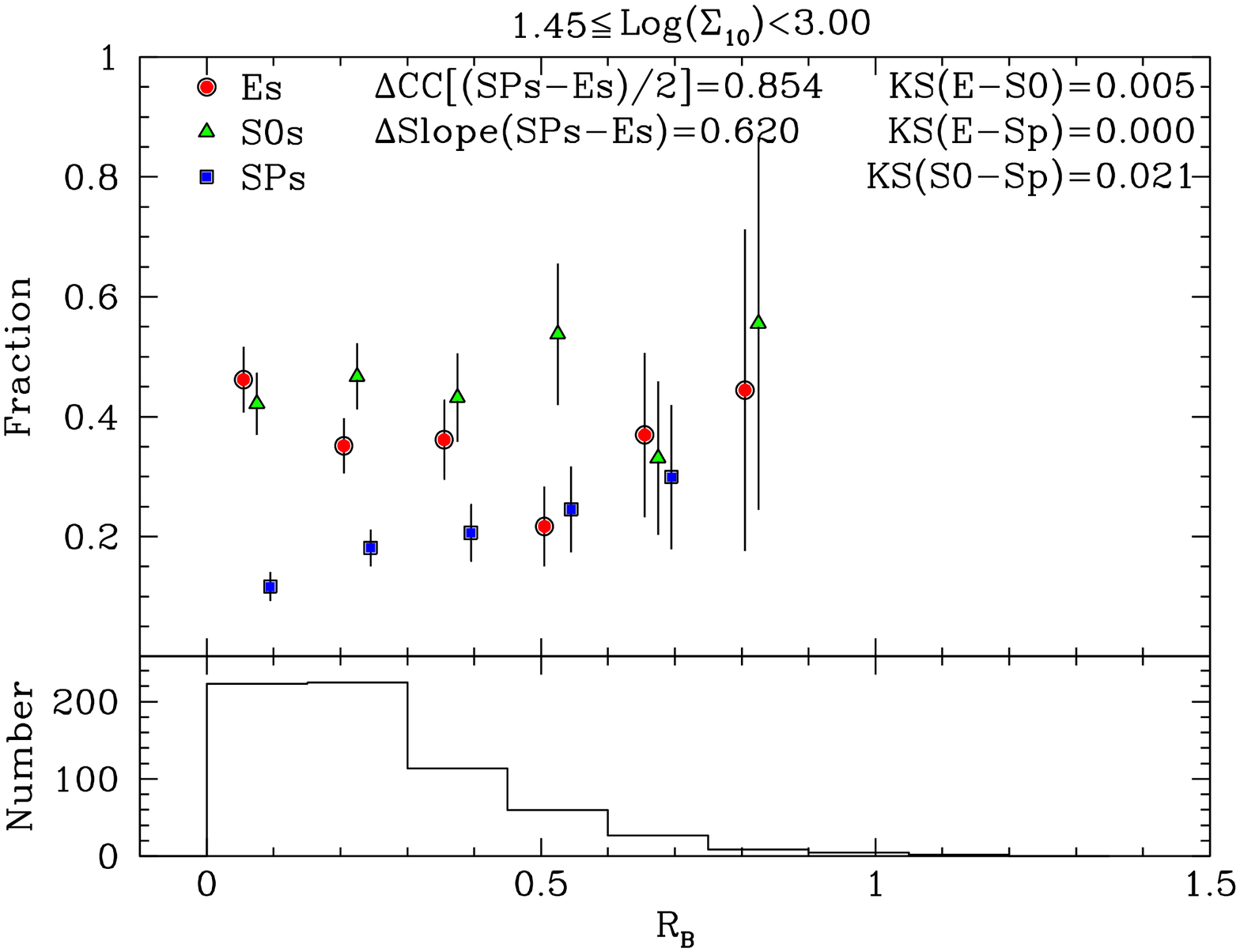}
\vspace{-0.5truecm}
\hspace{-0.8truecm}
\includegraphics[width=91mm]{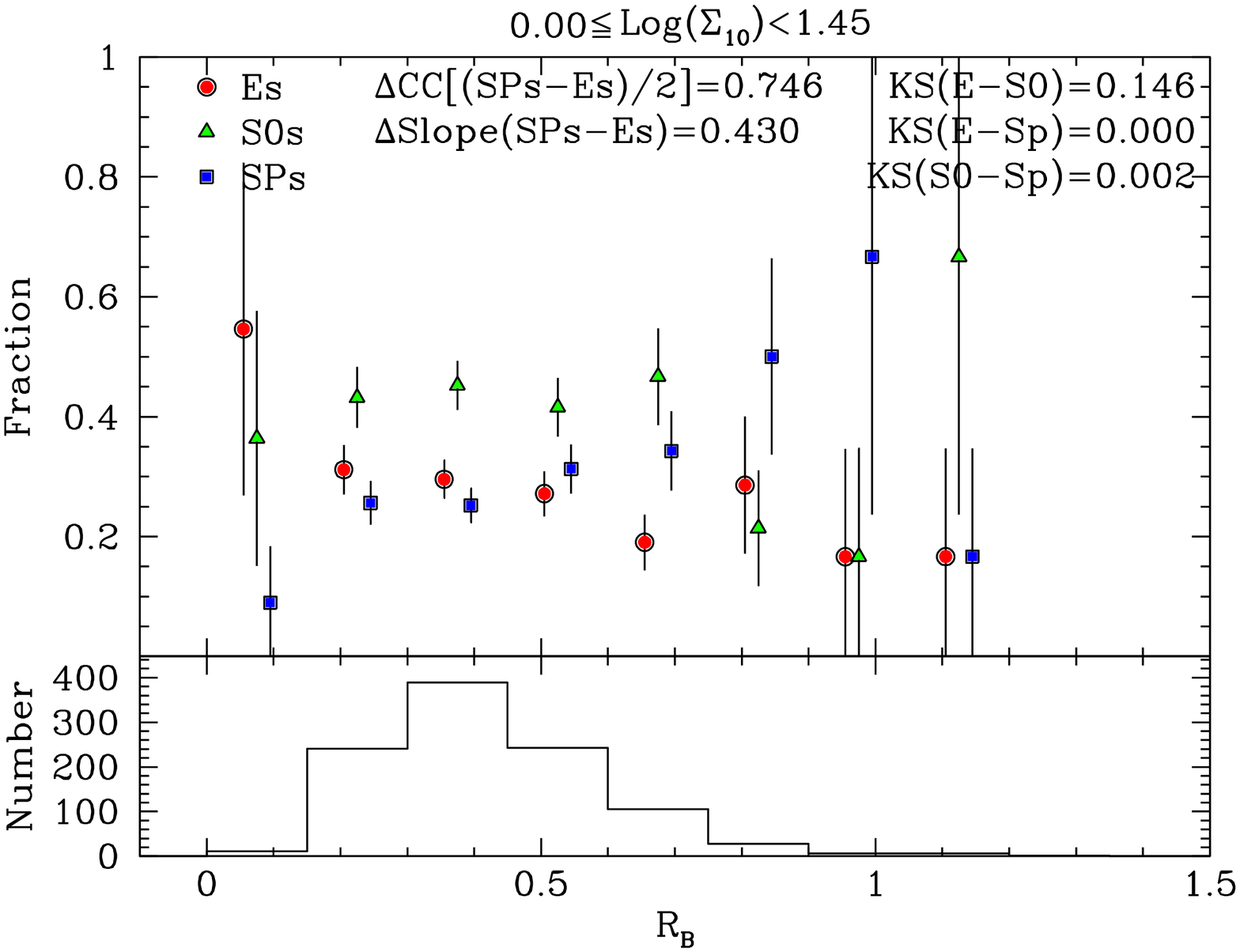}
\includegraphics[width=91mm]{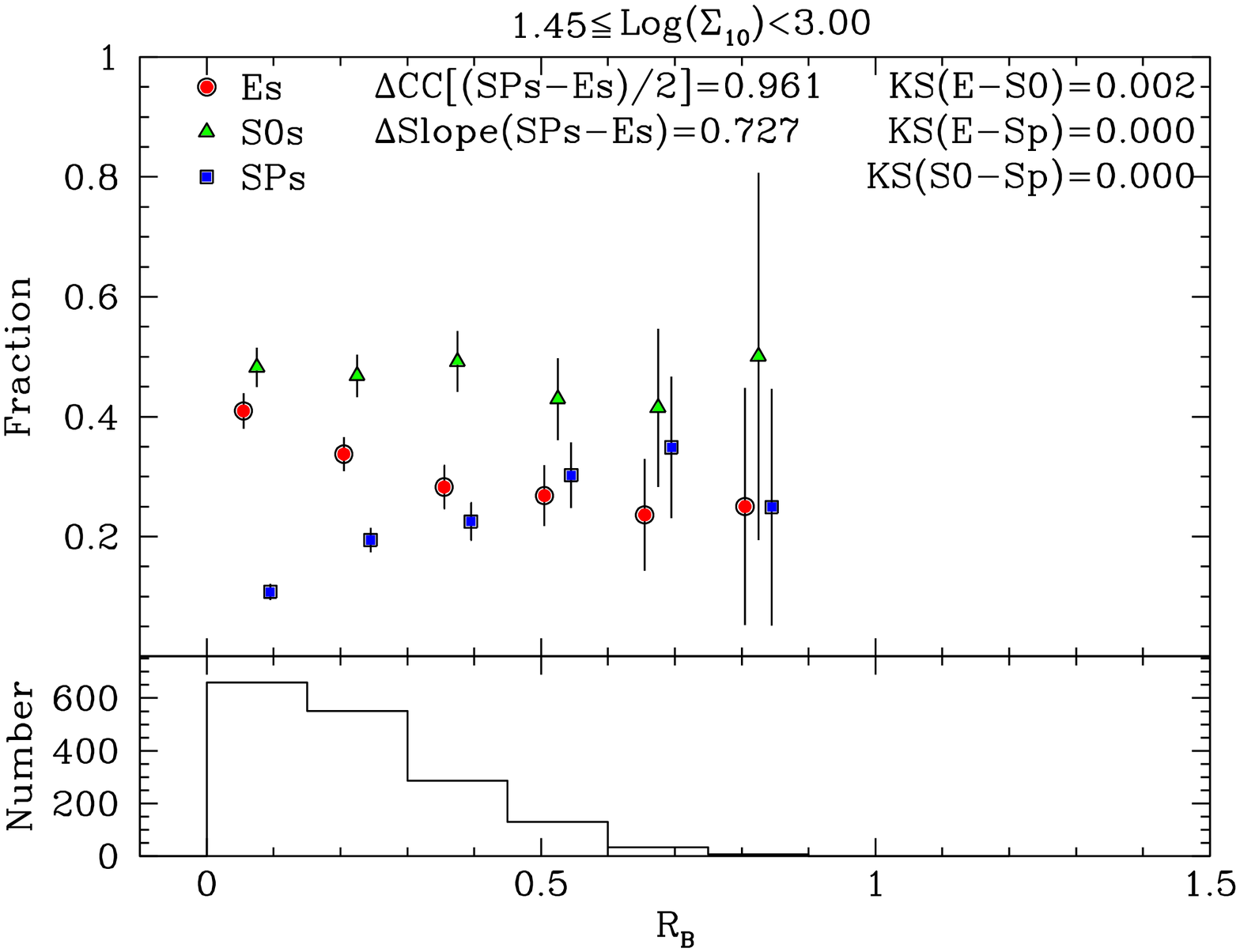}
\caption{$T$-$R$ relation for clusters with Log$\sigma<$2.85 (upper
  panels) and Log$\sigma\ge$2.85 (bottom panels) in different ranges of
  local density: Log$\Sigma_{10}<$1.45 (left panels) and
  1.45$\le$Log$\Sigma_{10}<$3 (right panels). Even at a first glance, in
  both intervals of Log$\sigma$ the $T$-$R$ relation appears to hold
  over the whole range of local density. The meaning of the symbols is
  as in Fig.~\ref{fig5}.}
\label{fignew2}
\end{figure*}

\begin{figure*}
\vspace{-0.5truecm}
\hspace{-0.8truecm}
\includegraphics[width=91mm]{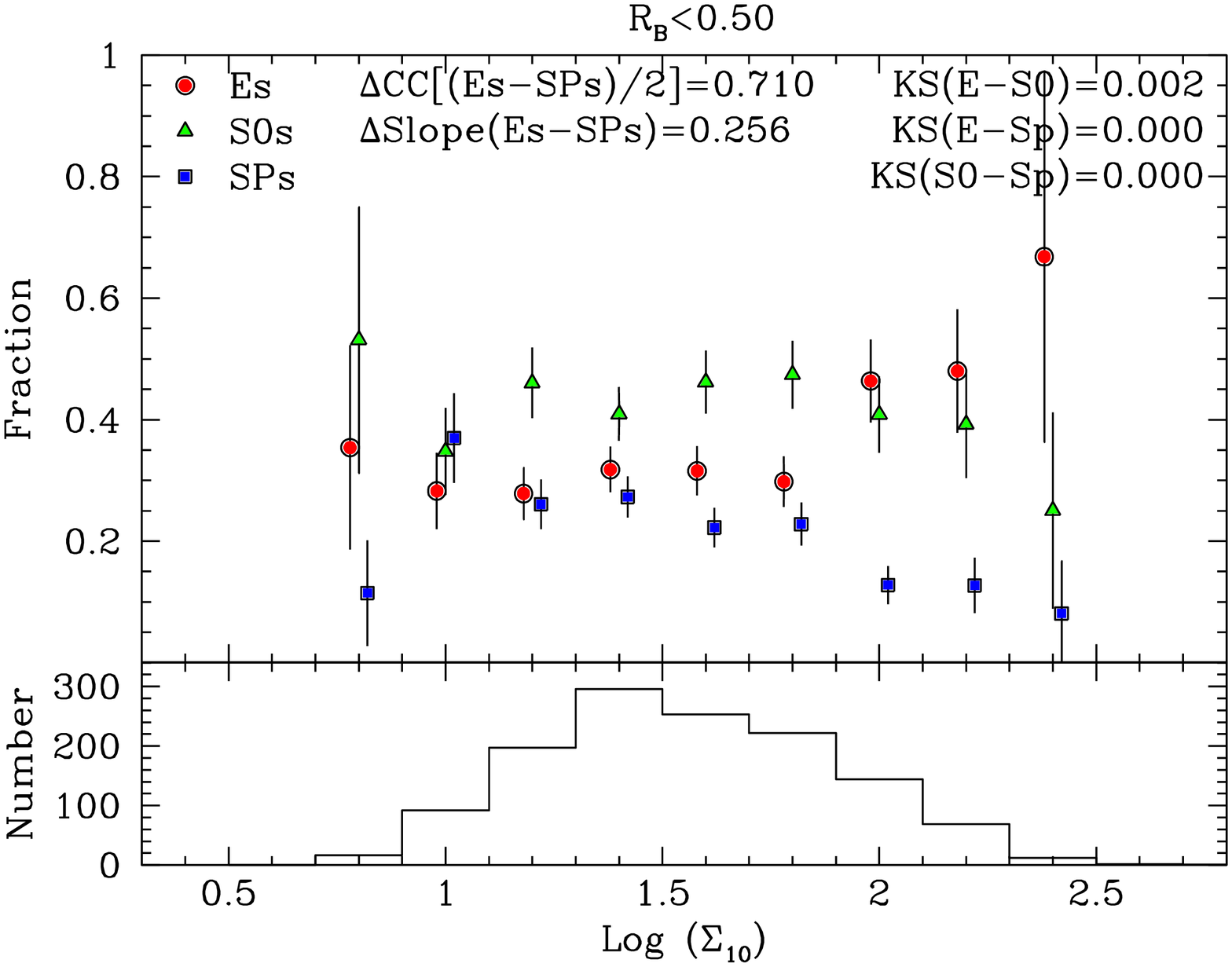}
\includegraphics[width=91mm]{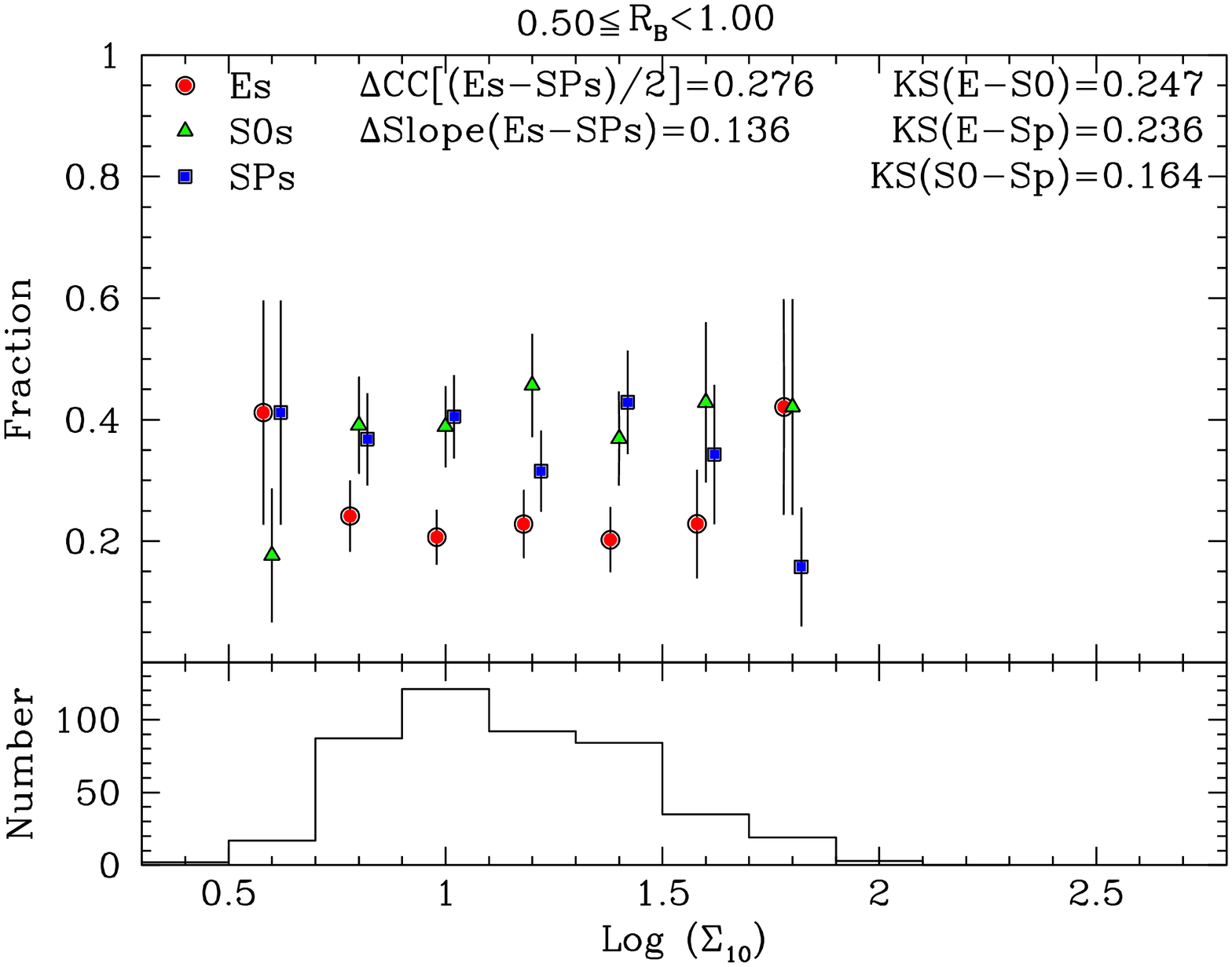}
\vspace{-0.5truecm}
\hspace{-0.8truecm}
\includegraphics[width=91mm]{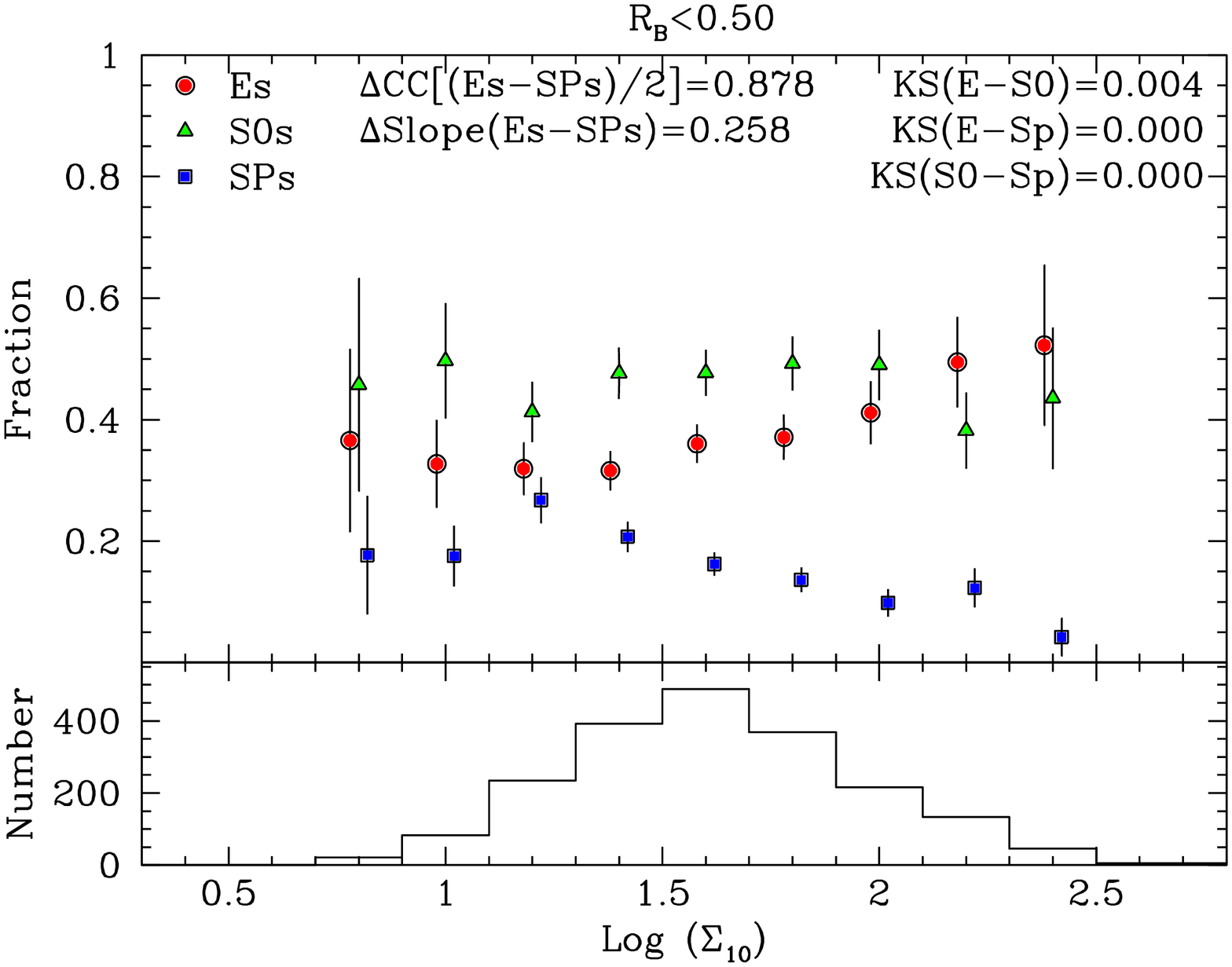}
\includegraphics[width=91mm]{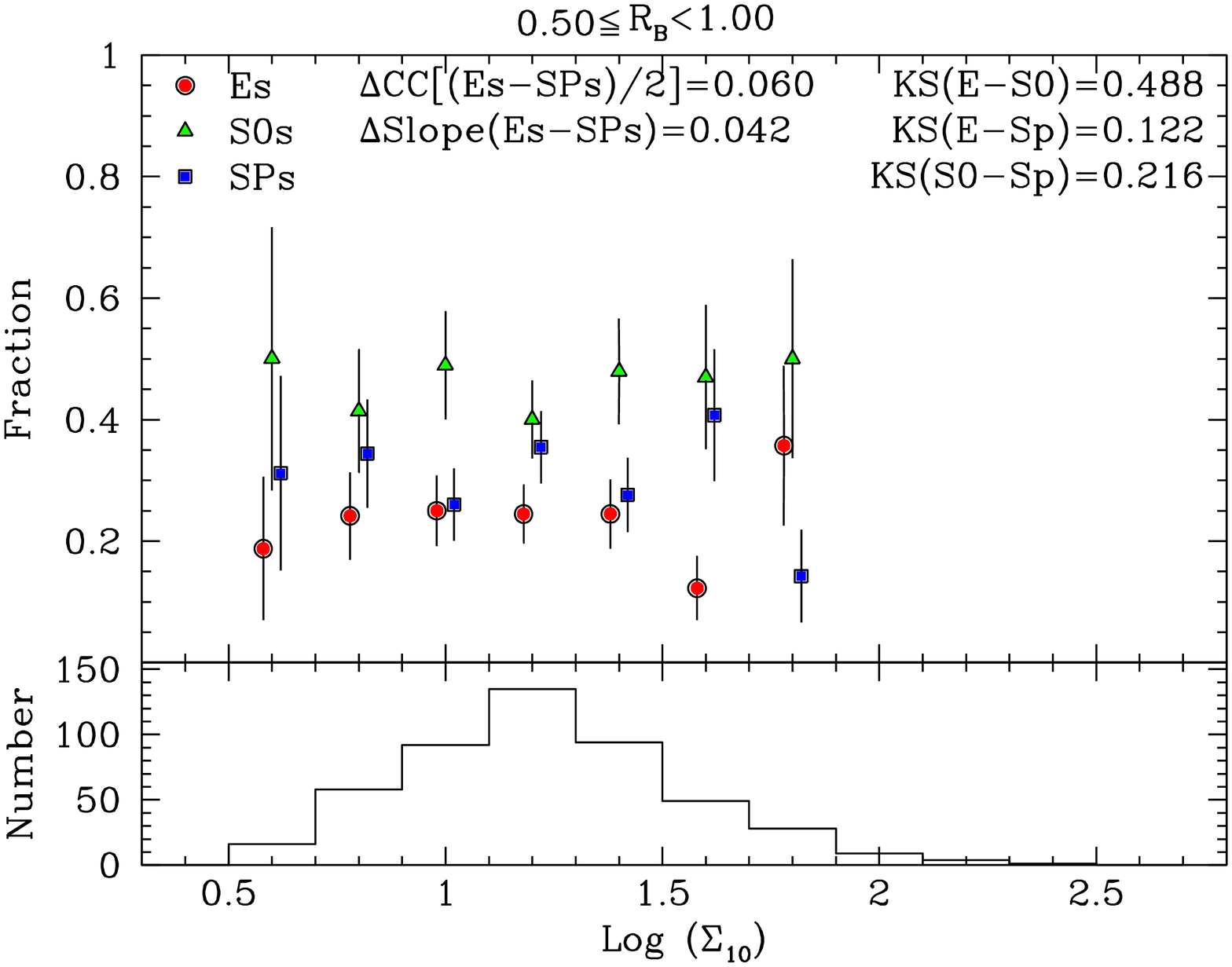}
\caption{$T$-$\Sigma$ relation for clusters with Log$L_X<$44.15 (upper
  panels) and Log$L_X\ge$44.15 (bottom panels) in different ranges of
  cluster-centric distance: $R_B<$0.5 (left panels) and
  0.5$\le R_B<$1 (right panels). Again, the
  $T$-$\Sigma$ relation appears to hold just in the inner cluster
  regions, irrespective of the cluster X-ray luminosity. The meaning
  of the symbols is as in Fig.~\ref{fig5}.}
\label{fignew1}
\end{figure*}

To include figures similar to previous ones for all 24 cases
reported in Tables ~\ref{tabsig} and \ref{tablx} would be a worthless
space waste. Thus we decided to show, in Figs~\ref{fignew2} and
\ref{fignew1}, just some examples illustrating the validity of the above
conclusions (see the figure captions for more details).

\begin{table*}
  \caption{Similar to Table~\ref{tabres}, but for two different intervals of Log$\sigma$}
\begin{tabular}{ccccccccc}
\hline
 & \multicolumn{4}{c}{$T$-$\Sigma$} & \multicolumn{4}{c}{$T$-$R$} \\
\hline
Log$\sigma$ &\ \ $R_B$ &  $\Delta$CC & $\Delta$Slope & KS(E-Sp)\ \ \  &\ \ \ Log$\Sigma_{10}$ & $\Delta$CC & $\Delta$Slope & KS(E-Sp) \\
\hline
  & All & 0.927 & 0.368 & 0.000 & 0--3 & 0.937 & 0.593 & 0.000 \\
  & $^{(2129)}$&$^{(0.034)}$&$^{(0.040)}$&&$^{(2112)}$&$^{(0.029)}$&$^{(0.054)}$&\\
 $<$2.85 & 0--0.5 & 0.757 & 0.277 & 0.000 & 0--1.45 & 0.871 & 0.527 & 0.000 \\
  & $^{(1222)}$&$^{(0.113)}$&$^{(0.064)}$&&$^{(1334)}$&$^{(0.054)}$&$^{(0.074)}$&\\
  & 0.5--1 & 0.289 & 0.100 & 0.425 & 1.45--3 & 0.854 & 0.620 & 0.000 \\
  & $^{(816)}$&$^{(0.226)}$&$^{(0.077)}$&&$^{(778)}$&$^{(0.096)}$&$^{(0.153)}$&\\
\hline
 & All & 0.862 & 0.298 & 0.000 & 0--3 & 0.936 & 0.690 & 0.000 \\
  & $^{(3106)}$&$^{(0.066)}$&$^{(0.042)}$&&$^{(3075)}$&$^{(0.031)}$&$^{(0.077)}$&\\
$\ge$2.85 & 0--0.5 & 0.812 & 0.280 & 0.000 & 0--1.45 & 0.746 & 0.430 & 0.000 \\
  & $^{(2591)}$&$^{(0.085)}$&$^{(0.054)}$&&$^{(1233)}$&$^{(0.113)}$&$^{(0.117)}$&\\
  & 0.5--1 & 0.083 & 0.075 & 0.083 & 1.45--3 & 0.961 & 0.727 & 0.000 \\
  & $^{(505)}$&$^{(0.192)}$&$^{(0.122)}$&&$^{(1842)}$&$^{(0.022)}$&$^{(0.073)}$&\\
\hline
\end{tabular}
\label{tabsig}
\end{table*}

\begin{table*}
  \caption{Similar to Table~\ref{tabres}, but for two different intervals of Log($L_X$)}
\begin{tabular}{ccccccccc}
\hline
 & \multicolumn{4}{c}{$T$-$\Sigma$} & \multicolumn{4}{c}{$T$-$R$} \\
\hline
Log($L_X$) &\ \ $R_B$ &  $\Delta$CC & $\Delta$Slope & KS(E-Sp)\ \ \  &\ \ \ Log$\Sigma_{10}$ & $\Delta$CC & $\Delta$Slope & KS(E-Sp) \\
\hline
  & All & 0.883 & 0.315 & 0.000 & 0--3 & 0.887 & 0.635 & 0.000 \\
  & $^{(2363)}$&$^{(0.054)}$&$^{(0.039)}$&&$^{(2192)}$&$^{(0.049)}$&$^{(0.077)}$&\\
 $<$44.15 & 0--0.5 & 0.710 & 0.256 & 0.000 & 0--1.45 & 0.852 & 0.550 & 0.000 \\
  & $^{(1524)}$&$^{(0.117)}$&$^{(0.068)}$&&$^{(1223)}$&$^{(0.062)}$&$^{(0.090)}$&\\
  & 0.5--1 & 0.276 & 0.136 & 0.236 & 1.45--3 & 0.885 & 0.949 & 0.000 \\
  & $^{(628)}$&$^{(0.223)}$&$^{(0.108)}$&&$^{(969)}$&$^{(0.075)}$&$^{(0.176)}$&\\
\hline
 & All & 0.943 & 0.297 & 0.000 & 0--3 & 0.936 & 0.615 & 0.000 \\
  & $^{(3041)}$&$^{(0.025)}$&$^{(0.027)}$&&$^{(2895)}$&$^{(0.031)}$&$^{(0.067)}$&\\
$\ge$44.15 & 0--0.5 & 0.878 & 0.258 & 0.000 & 0--1.45 & 0.740 & 0.432 & 0.000 \\
  & $^{(2228)}$&$^{(0.054)}$&$^{(0.038)}$&&$^{(1277)}$&$^{(0.116)}$&$^{(0.107)}$&\\
  & 0.5--1 & 0.060 & 0.042 & 0.122 & 1.45--3 & 0.944 & 0.562 & 0.000 \\
  & $^{(653)}$&$^{(0.225)}$&$^{(0.120)}$&&$^{(1618)}$&$^{(0.033)}$&$^{(0.070)}$&\\
\hline
\end{tabular}
\label{tablx}
\end{table*}

\subsubsection{Subclustering}\label{Sec6c2}

\begin{figure}
\vspace{-0.5truecm}
\includegraphics[width=91mm]{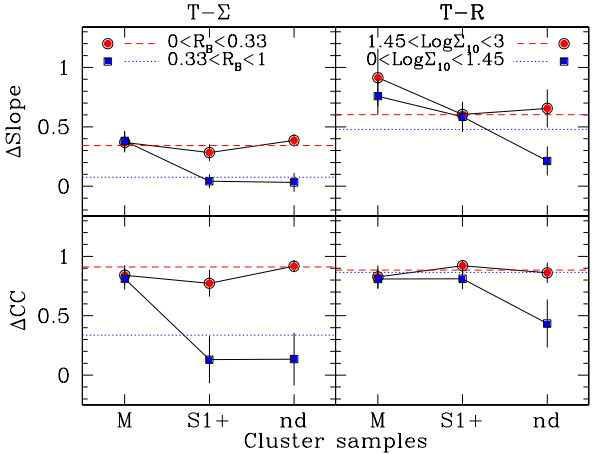}
\vspace{-0.5truecm}
\caption{Coefficients $\Delta$Slope (upper panels) and $\Delta$CC
  (lower panels) of the $T$-$\Sigma$ (left panels) and $T$-$R$ (right
  panels) relations for the three cluster subsamples ( {\bf M}, {\bf
    S1+} and {\bf nd}) defined in Section~\ref{Sec6c2}. The full dots
  and squares (red and blue, in the electronic version) refer to
  different ranges of $R_B$ and $Log\Sigma_{10}$, for the $T$-$\Sigma$
  and $T$-$R$ relations, respectively.  The dashed and dotted lines
  (again red and blue, in the electronic version) correspond to the
  values found in the different cases for the whole sample of WINGS
  clusters.}
\label{fig9}
\end{figure}

 The $T$-$\Sigma$ relation has been found by D80 to hold for both
regular (``relaxed'') and irregular (``non relaxed'') clusters of the
local Universe and this finding has given support to the idea that it
is a sort of general rule in the realm of galaxies. Indeed, since, by
definition, the $T$-$\Sigma$ relation concerns the local environment,
assuming such a relation to have a general validity implies it should
be found everywhere, no matter which is the global structure of the
cluster. On the contrary, the lack of $T$-$\Sigma$ relation in
irregular (clumpy) clusters would suggest that either it is just a
by-product of the $T$-$R$ relation (assuming it ever holds in such
kind of clusters), or that in the cluster substructures (infalling
groups?) the morphological segregation and processing still is
on-going or to come, as suggested by D97 for intermediate redshift
clusters (a rather unlikely possibility for nearby clusters,
indeed). Therefore, it is of some interest to test, in our large
sample of nearby clusters, the effects of subclustering on the
$T$-$\Sigma$ and $T$-$R$ relations. To this aim, following the
classification given by the catalog of substructures in WINGS clusters
\citep[][see Table A1 therein]{rame07}, we have divided our cluster
sample in three sub-samples: {\it (i)} the sample {\bf M} includes the
15 WINGS clusters for which just the main structure (M) was detected
by the DEDICA algorithm \citep{pisa93,pisa96}; {\it (ii)} the sample
{\bf S1+} includes the 40 clusters for which DEDICA found at least one
substructure (S) at the same redshift of the main structure; {\it
  (iii)} the sample {\bf nd} (not detected) contains the 22 WINGS
clusters whose irregular and very clumpy structure prevented DEDICA
from detecting any significant structure inside them. For these
cluster samples, Figure~\ref{fig9} reports the coefficients
$\Delta$Slope (upper panels) and $\Delta$CC (lower panels) of the
$T$-$\Sigma$ (left panels) and $T$-$R$ (right panels) relations, in
two ranges of $R_B$ and $Log\Sigma_{10}$, respectively. The dashed and
dotted lines (red and blue, in the electronic version) correspond to
the values found for the whole sample of WINGS clusters in the
different cases.

Some interesting things come to light from Figure~\ref{fig9}: {\it
  (a)} in the inner part of clusters ($R_B<$0.33) and in the regions
of high local density (Log$\Sigma_{10}>$1.45), the $T$-$\Sigma$ and
$T$-$R$ relations (respectively) turn out to be quite strong in all
three sub-samples (full dots in the figure; red in the electronic
version). In particular, in each case the strength of the relations is
similar to that of the corresponding one relative to the whole cluster
sample, no matter which is the subclustering level
(dashed lines; see also Table~\ref{tabres}); {\it
  (b)} for $R_B>$0.33, the relation $T$-$\Sigma$ turns out to be as
strong as in the inner part of clusters just for the subsample {\bf
  M}, which presumably contains the most regular (``relaxed'')
clusters, while for the other two sub-samples ({\bf S1+} and {\bf nd})
the lack of relation is confirmed at a significance level even greater
than in the case of the whole cluster sample; {\it (c)} in the lower
bin of local density (Log$\Sigma_{10}<$1.45), the $T$-$R$ relation
remains strong for both the {\bf M} and {\bf S1+} sub-samples,
becoming weaker just for the sub-sample {\bf nd}.

\begin{figure*}
\vspace{-0.5truecm}
\hspace{-0.8truecm}
\includegraphics[width=91mm]{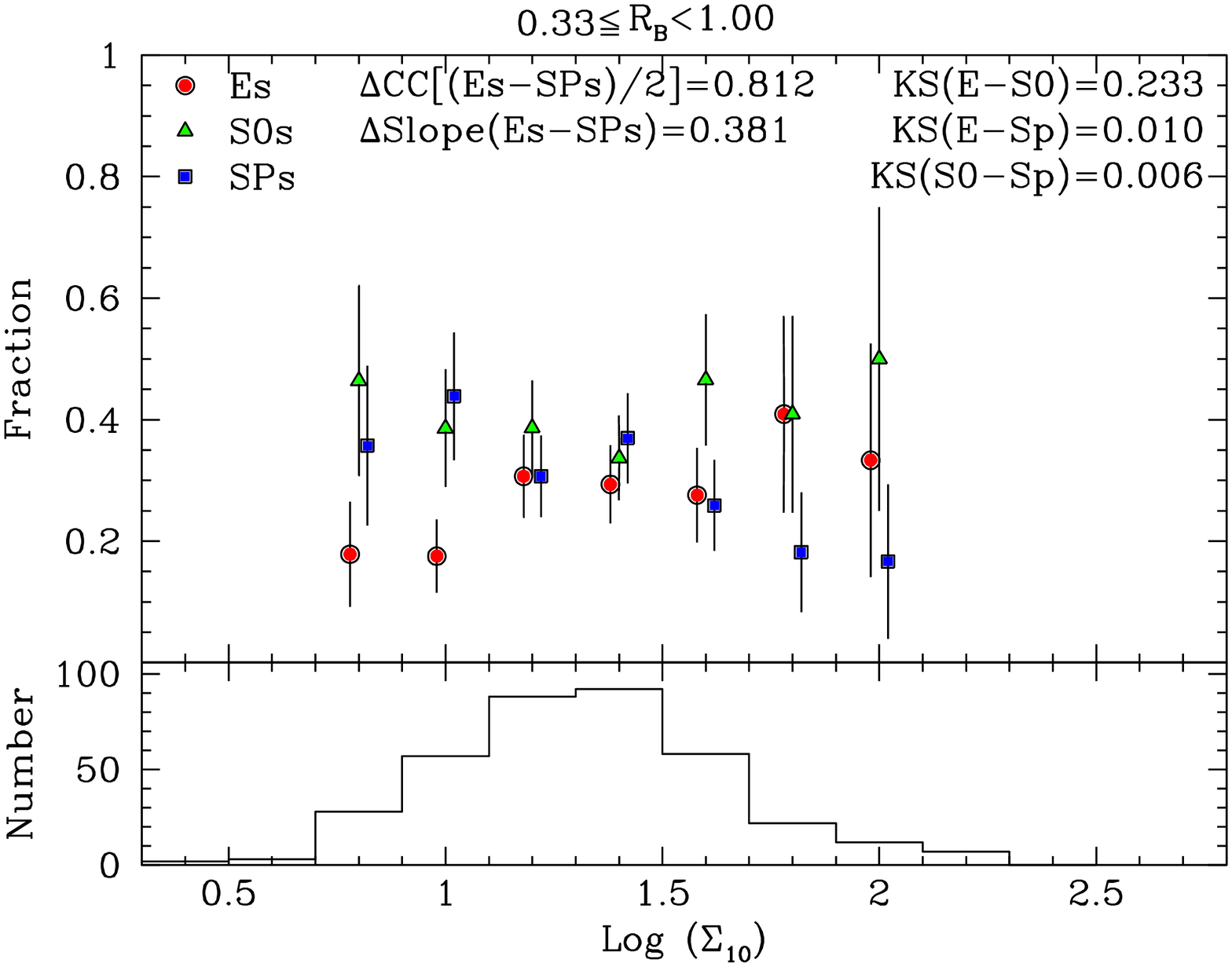}
\includegraphics[width=91mm]{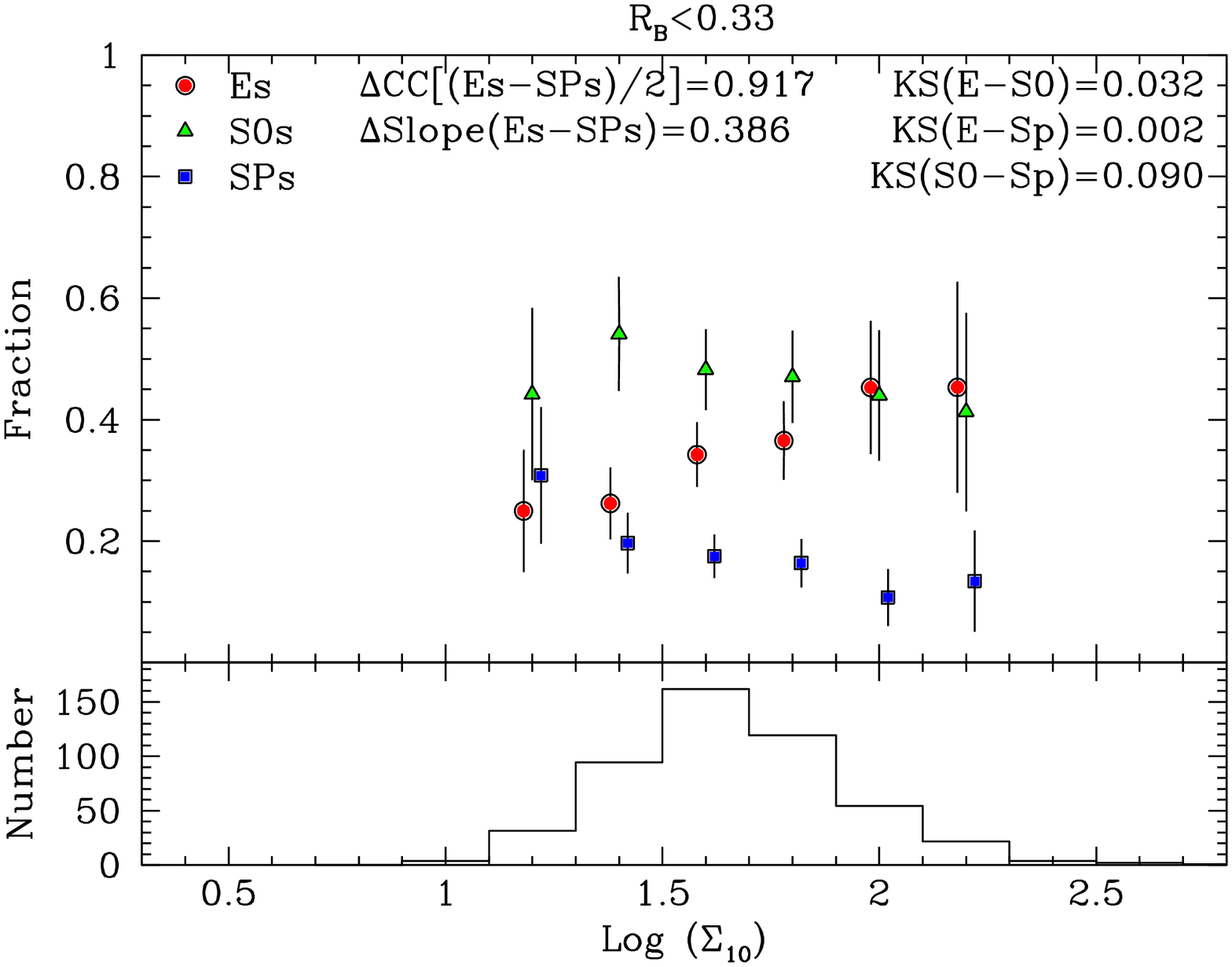}
\vspace{-0.5truecm}
\hspace{-0.8truecm}
\includegraphics[width=91mm]{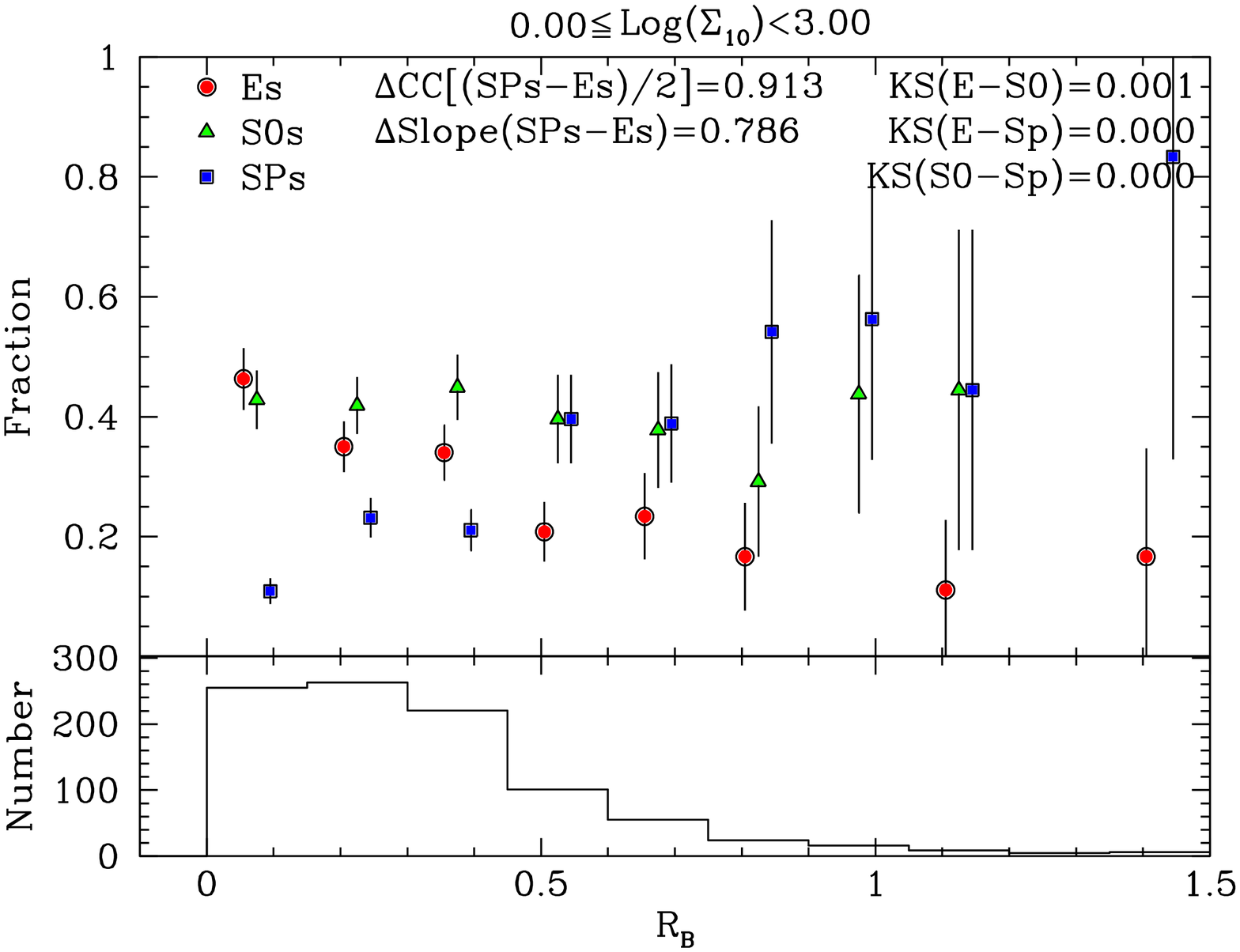}
\includegraphics[width=91mm]{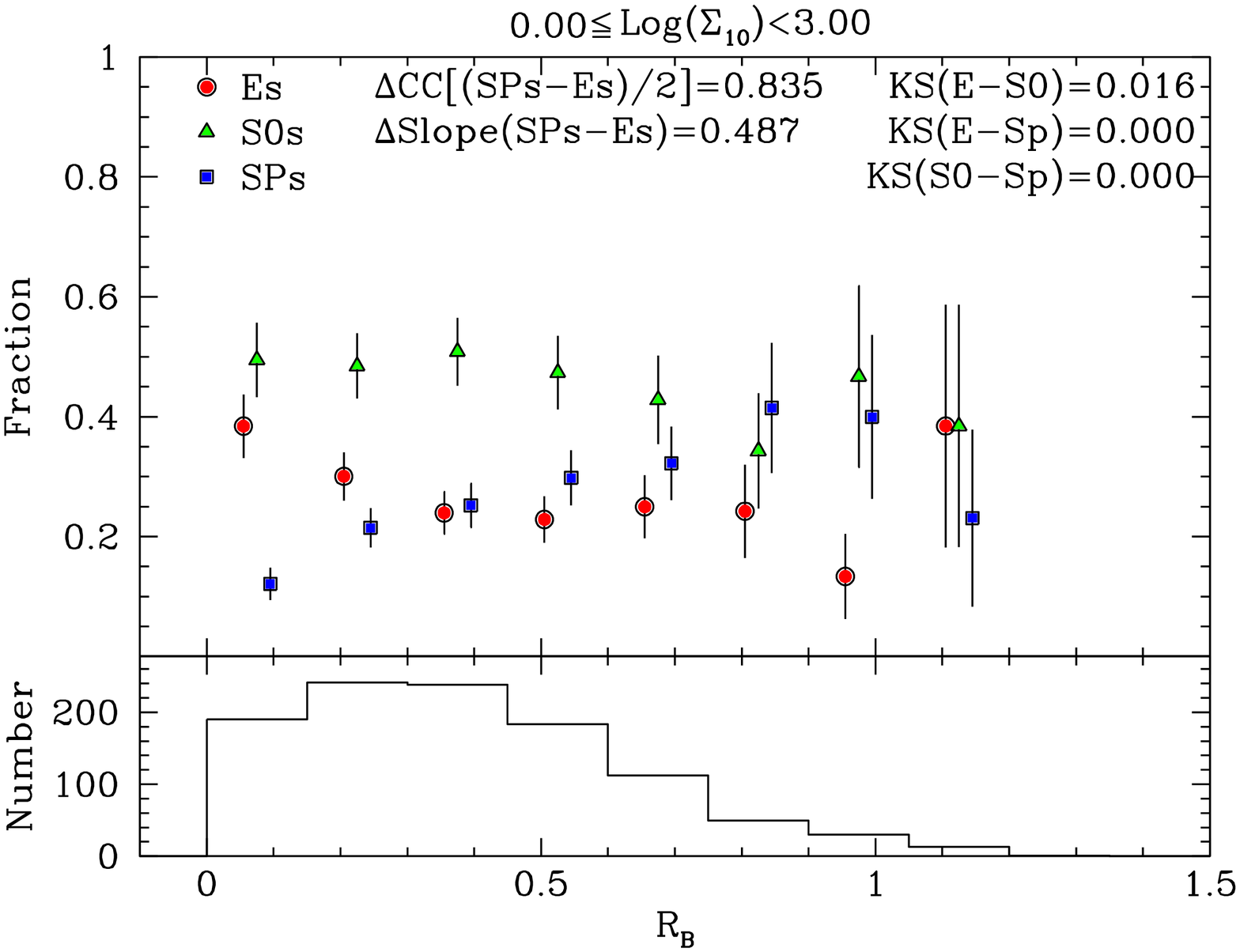}
\caption{$T$-$\Sigma$ and $T$-$R$ relations (upper and bottom panels,
  respectively) for regular (left panels; sample {\bf M}) and very
  irregular (right panels; sample {\bf nd}) clusters. The $R_B$ and
  $\Sigma_{10}$ intervals tried out in each case are indicated on the
  top of each panel.  The meaning of the symbols is
  as in Fig.~\ref{fig5}.}
\label{fignew3}
\end{figure*}

In Figure~\ref{fignew3} the $T$-$\Sigma$ and $T$-$R$ relations (upper
and bottom panels, respectively) for the regular (left panels; sample
{\bf M}) and very irregular (right panels; sample {\bf nd}) clusters
are illustrated for some particularly interesting cases. The upper
panels clearly show that: {\it (i)} for regular clusters the
$T$-$\Sigma$ relation holds outside the very inner regions too (at
variance with our finding relative to the global cluster sample); {\it
  (ii)} in the inner cluster regions the $T$-$\Sigma$ relation holds
even for very irregular clusters (not a trivial thing, indeed). The
bottom panels show that the $T$-$R$ relation, in the whole range of
$\Sigma_{10}$, holds for both regular and very irregular clusters.

\subsection{$T$-$\Sigma$ and $T$-$R$ for different stellar masses}\label{Sec6d}

\citet{peng10}, using the surveys SDSS and zCOSMOS
\citep{scov07,lilly07}, have analysed the fraction of quenched
galaxies in the general field as a function of both the stellar mass
and the environment, over the cosmic time. More closely in connection
with our investigation, the dependence of global morphological
fractions on galaxy stellar mass in clusters at different redshifts
has been analysed by \citet{vulc11}.  However, the possible dependence
of the $T$-$\Sigma$ and $T$-$R$ relations on $M_*$ has not been
investigated so far, apart from some particular aspects of the problem
analysed by \citet{bamf09}.  In this analysis the fraction of
early-type galaxies is found to increase with the local density in
each bin of stellar mass, the slope of the Fraction--Log$\Sigma_{10}$
relation being almost the same in all bins. However, this study is
based on the morphological classifications provided by the Galaxy Zoo
project \citep{lint11}, being thus limited by the lack of distiction
between ellipticals and S0 galaxies.

Stellar masses of WINGS galaxies \citep{frit11} have been determined
by fitting the optical spectrum (in the range
$\sim$3600--$\sim$7000$\AA$) with the spectro-photometric model fully
described in \citet{frit07}. In this model, all the main
spectro--photometric features are reproduced by summing the
theoretical spectra of Simple Stellar Population (SSP) of 13 different
ages (from 3$\times$10$^6$ to $\sim$14$\times$10$^9$ years) and
assuming a Salpeter IMF. Dust extinction is allowed to vary as a
function of SSP age and the metallicity can vary among three values:
Z=0.004, Z=0.02 and Z=0.05. These mass estimates were then corrected
for color gradients within each galaxy (i.e. for the difference in
color within the fibre and over a 10kpc diameter) and were converted
to the \citet{krou01} IMF. \citet{frit11} provide mass estimates for
$\sim$5,300 WINGS galaxies, 1,540 of them belonging to the galaxy
sample we use in this paper. In \citet{vulc11} it is shown that these
mass estimates are in fairly good agreement with those derived
using the rest-frame $(B-V)$ colors and the recipe given in \citet{bdj01}.

\begin{table*}
  \caption{Similar to Table~\ref{tabres}, but for two different intervals of Log($M_*$)}
\begin{tabular}{ccccccccc}
\hline
 & \multicolumn{4}{c}{$T$-$\Sigma$} & \multicolumn{4}{c}{$T$-$R$} \\
\hline
Log($M_*$) &\ \ $R_B$ &  $\Delta$CC & $\Delta$Slope & KS(E-Sp)\ \ \  &\ \ \ Log$\Sigma_{10}$ & $\Delta$CC & $\Delta$Slope & KS(E-Sp) \\
\hline
  & All & 0.508 & 0.195 & 0.001 & 0--3 & 0.697 & 0.510 & 0.000 \\
  & $^{(716)}$&$^{(0.185)}$&$^{(0.096)}$&&$^{(716)}$&$^{(0.149)}$&$^{(0.149)}$&\\
 $<$10.76 & 0--0.5 & 0.679 & 0.290 & 0.009 & 0--1.45 & 0.467 & 0.404 & 0.006 \\
  & $^{(523)}$&$^{(0.150)}$&$^{(0.103)}$&&$^{(365)}$&$^{(0.213)}$&$^{(0.240)}$&\\
  & 0.5--1 & -0.020 & -0.057 & 0.168 & 1.45--3 & 0.638 & 0.554 & 0.002 \\
  & $^{(182)}$&$^{(0.337)}$&$^{(0.327)}$&&$^{(351)}$&$^{(0.197)}$&$^{(0.210)}$&\\
\hline
  & All & 0.879 & 0.354 & 0.000 & 0--3 & 0.876 & 0.681 & 0.000 \\
  & $^{(824)}$&$^{(0.055)}$&$^{(0.053)}$&&$^{(809)}$&$^{(0.071)}$&$^{(0.118)}$&\\
 $>$10.76 & 0--0.5 & 0.738 & 0.280 & 0.004 & 0--1.45 & 0.753 & 0.644 & 0.001 \\
  & $^{(590)}$&$^{(0.120)}$&$^{(0.081)}$&&$^{(395)}$&$^{(0.134)}$&$^{(0.207)}$&\\
  & 0.5--1 & -0.183 & -0.096 & 0.404 & 1.45--3 & 0.772 & 0.637 & 0.001 \\
  & $^{(225)}$&$^{(0.289)}$&$^{(0.186)}$&&$^{(414)}$&$^{(0.142)}$&$^{(0.190)}$&\\
\hline
\end{tabular}
\label{tabmass}
\end{table*}

\begin{figure}
\vspace{-0.5truecm}
\includegraphics[width=91mm]{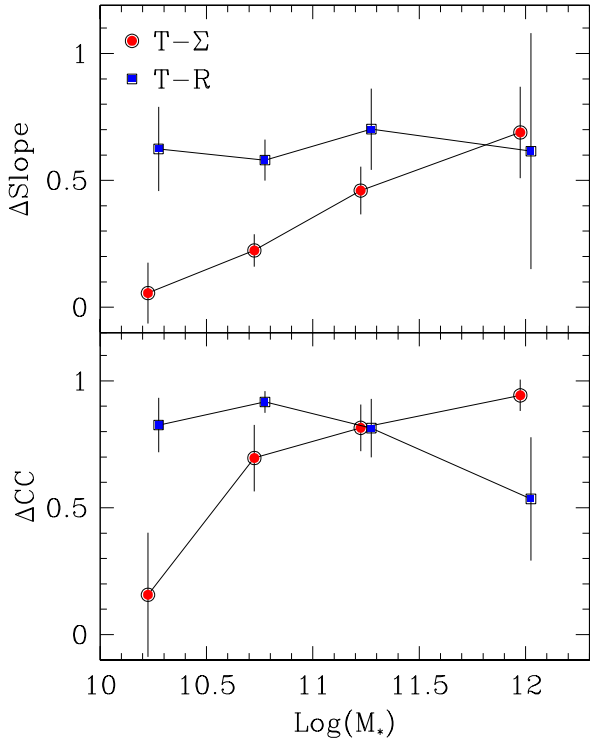}
\vspace{-0.5truecm}
\caption{$\Delta$Slope (upper panel) and $\Delta$CC (lower panel) as a
  function of Log($M_*$) for both $T$-$\Sigma$ (full dots; red in the
  electronic version) and $T$-$R$ (full squares; blue in the electronic
  version)}
\label{fig10}
\end{figure}

Table~\ref{tabmass} is similar to previous Tables~\ref{tabsig} and
\ref{tablx}, but it splits the WINGS galaxies in two intervals of
Log($M_*$). In this case the splitting value roughly coincides with
the median of the Log($M_*$) distribution.  Provided that the
$T$-$\Sigma$ for this reduced galaxy sample proves again itself to be
weak or absent outside the inner cluster regions, from
Table~\ref{tabmass} it turns out that the strenght of the $T$-$\Sigma$
relation is greater in the high-mass than in the low-mass bin. We
decided to further explore this point dividing the stellar mass
interval in four relatively narrow bins of size 0.5, starting from
Log($M_*$)=10~~\footnote{The spectroscopic magnitude limit of the
  WINGS survey is V=20, corresponding to a mass limit of
  Log($M_*$)=9.8. \citet{vulc11} actually report that the completeness
  of mass limited samples in WINGS is reached for
  Log($M_*$)=10.5. Thus, the lowest mass bin in Fig.~\ref{fig10} could
  still suffer from some bias.}. Figure~\ref{fig10} illustrates the
results. It clearly shows that, while for the $T$-$R$ relation both
$\Delta$Slope and $\Delta$CC turn out to be almost stable in the four
bins of Log($M_*$), they are strongly increasing functions of the
galaxy stellar mass for the $T$-$\Sigma$ relation. In particular, in
the lowest mass bin, both $\Delta$Slope and $\Delta$CC indicate that
the $T$-$\Sigma$ is almost absent.

We have also tried to account for spectroscopic incompleteness by
applying a statistical correction to our sample. This is obtained by
weighting each galaxy by the inverse of the ratio of the number of
spectra yielding a redshift to the total number of galaxies in the
photometric catalogue, in bins of 1~mag \citep{cava09}. We do not
report here the results of such additional analysis. We just mention
that they turn out to be quite similar to those shown in
Table~\ref{tabmass} and in Figure~\ref{fig10}.

This might induce to draw the conclusion that the lack of $T$-$\Sigma$ we
observe outside the inner cluster regions is due to an excess in these
regions of galaxies in the lowest stellar mass bin with respect to the
other bins. However, this conclusion turns out to be ruled out
by the 2S-KS applied to the clustercentric distances of galaxies in
the four mass bins.

It is well known \citep[see for instance][]{vulc11} that the three
broad morphological types (E/S0/Sp) have quite different stellar mass
distributions. This is confirmed by the 2S-KS applied to our galaxy
sample. It might be speculated that this fact, combined with the
dependence of the mass distribution on the local density
\citep{vulc12}, is fully responsible of the very existence of the
$T$-$\Sigma$ relation. In other words, the $T$-$\Sigma$ could just be
a consequence of the combined dependence of both the morphology and
the local density on the stellar mass. This hypothesis would obviously
imply that the $T$-$\Sigma$ relation should not be observed at any
(fixed) galaxy mass.

\begin{figure}
\vspace{-0.5truecm}
\includegraphics[width=100mm]{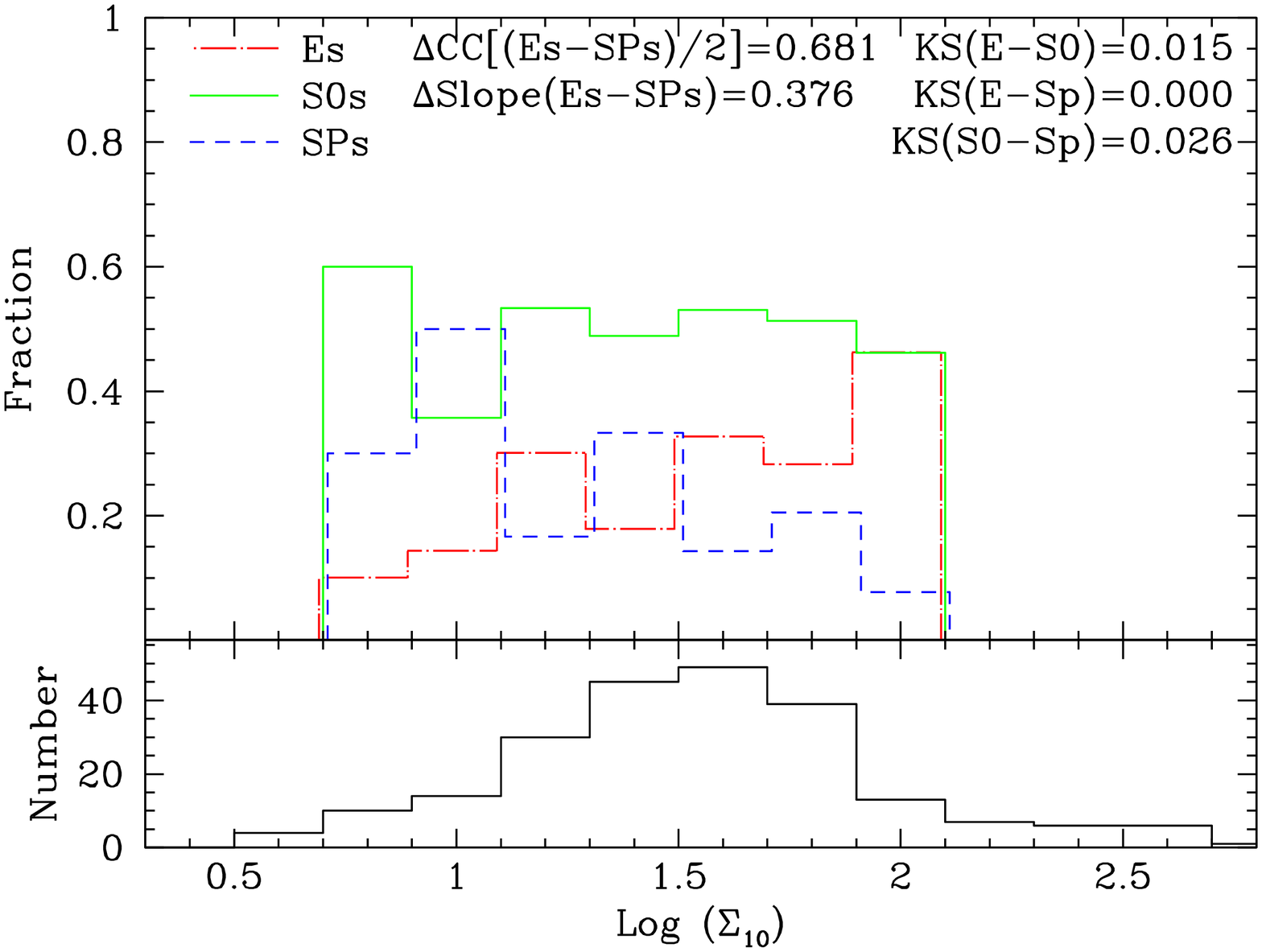}
\includegraphics[width=100mm]{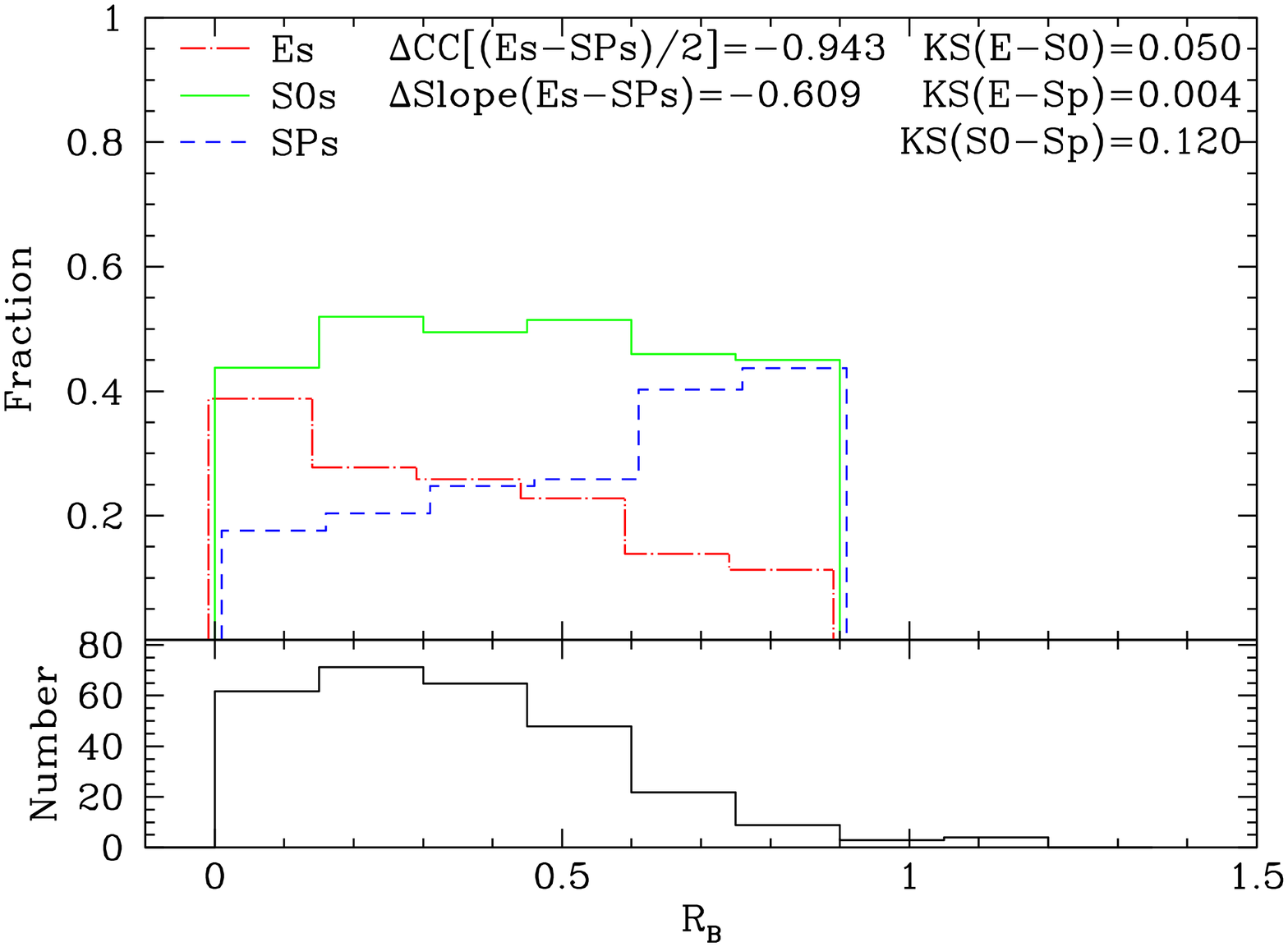}
\caption{$T$-$\Sigma$ (upper panel) and $T$-$R$ (lower panel) for galaxies in the
  Log($M_*$) bins (11--11.3) and (10.3--10.6), respectively.}
\label{fig11}
\end{figure}

The upper panel of Figure~\ref{fig11} contradicts such an expectation.
It shows that, in spite of the poor statistics (218 galaxies), the $T$-$\Sigma$
turns out to be remarkably strong even in quite narrow stellar mass
bins [11$<$Log($M_*$)$<$11.3]. For the sake of clarity, in this plot
we use the histograms to represent the $T$-$\Sigma$, instead of the points,
since the big error bars associated to the points whould weight down
too much the figure. It is also worth noticing that we have chosen the
narrow mass bin in the high-mass part of the mass distribution, since
the $T$-$\Sigma$ progressively weakens towards low-mass galaxies (see
Figure~\ref{fig10}). The lower panel of Figure~\ref{fig11} shows that,
similarly to $T$-$\Sigma$, the $T$-$R$ relation appears to be strong enough also in a
narrow stellar mass bin, while, at variance with the $T$-$\Sigma$ (and
according to Figure~\ref{fig10}), it is quite strong also for galaxies
in the low-mass region of the distribution
[10.3$<$Log($M_*$)$<$10.6].

\section{Summary and open issues}\label{Sec7}

In the present analysis, performed using the WINGS database of galaxies
in nearby clusters, we have shown that:

\begin{itemize}
\item the correlation between morphological fractions and local
  density ($T$-$\Sigma$) exists only in the very inner regions of
  nearby clusters, almost vanishing outside 1/3$\times R_{200}$, apart
  from very regular (non substructured) clusters, for which the
  $T$-$\Sigma$ relation also holds outside the inner regions;
\item in contrast, the strenght of the correlation between
  morphological fractions and clustercentric distance ($T$-$R$)
  remains almost unchanged over the whole range of local density for
  both regular and irregular clusters, only slightly lessening for
  extremely clumpy clusters (bad center determination?);
\item a couple of suitable tests and two different sets of numerical
  simulations support the validity of these results even considering
  the possible biases arising from the limited cluster area coverage
  of the WINGS imaging;
\item the above findings hold irrespective of both the global
  cluster mass (velocity dispersion and X-ray emission) and the
  stellar mass of galaxies;
\item the strength of the $T$-$\Sigma$ relation (where present)
  increases with increasing galaxy stellar mass, while
  this effect is not found for the $T$-$R$ relation;
\item both the $T$-$\Sigma$ (where present) and the $T$-$R$ relations
  are remarkably strong even in quite narrow stellar mass
  bins. In particular, for the $T$-$\Sigma$ relation, this rules out
  the hypothesis that it could just be a consequence of the combined
  dependence of both morphology and local density on
  stellar mass.
\end{itemize}

These results could lead us to conclude that the parameter actually
driving morphological fractions in nearby clusters is the distance
from the cluster center, rather than the local density, as commonly
believed. In this scenario the $T$-$\Sigma$ relation would just be a
by-product of the $T$-$R$ relation, as already claimed by
\citet{WG93}. Against this conclusion, however, one can bring forward
the argument (challenged by \citealt{W95}) that the $T$-$\Sigma$
relation is also found in the general field
\citep[groups$+$pairs$+$single
galaxies;][]{bhav81,desou82,post84,hels03}.  In fact, according to the
standard paradigm, galaxy clusters are progressively built up through
continuous infalling into the main cluster halo of galaxies coming
from the surrounding field.  Thus, it is worth asking oneself whether
the lack of the $T$-$\Sigma$ in the out-of-center part of clusters
really implies the fall of the $T$-$\Sigma$ paradigm or, instead, some
other mechanism should be invoked to interpret our findings.

First, we note that the results illustrated in this paper
(Section~\ref{Sec6c2}, in particular) suggest that the $T$-$\Sigma$
relation holds only in dynamically evolved regions of nearby clusters,
{\it i.e.} the whole clusters, or just their inner parts, for regular
(``relaxed'') or substructured (``non relaxed'') clusters,
respectively. Therefore, it seems to us that the apt question is: what
is causing the observed $T$-$\Sigma$ weakening in dynamically ``non
relaxed'' regions? We speculate that some mechanism of morphological
broadening/redistribution is responsible for the $T$-$\Sigma$
weakening in the intermediate regions of substructured (``non
relaxed'') clusters.

Two different (and perhaps complementary) ways to attain this effect
are conceivable: {\it (i)} general field galaxies lose (release) the
dependence of their morphology on the local density when they infall
into the cluster, recovering it only when (and where) the dynamical
equilibrium has been already reached (only inner regions for
substructured, ``non relaxed'' clusters); {\it (ii)} galaxies having
already experienced the very dense interiors of clusters and moving
back to the intermediate/outer regions, or 'free' galaxies coming in
these regions from different sub-halos are `$T$-$\Sigma$ untied', thus
diluting the relation likely existing in the infalling galaxy
population.

The first mechanism could result from reshuffling of galaxy locations
due the gravitational interactions within the cluster.  The second
mechanism envisages a scenario in which galaxies responsible for
morphological broadening could come from the innermost cluster
regions (likely after one or more passages close to the cluster
center) and/or could be detached from their original infalling halos,
therefore losing their pristine correlation between morphology and
$\Sigma$.

Of course, the two outlined mechanisms of morphological broadening
could operate in tandem. Unfortunately, there are no observational
hints about which one of them is prevailing, or even about their very
existence. As a matter of fact, it is presently unknown which process
is responsible for the weakening of $T$-$\Sigma$ in the intermediate
cluster regions. The increasingly powerful and sophisticated
$\Lambda CDM$ models of galaxy formation and evolution inside DM halos
could help shed some light on these findings.

\section*{Acknowledgments}

{We acknowledge partial financial support by contract PRIN/MIUR 2009:
  “Dynamics and Stellar Populations of Superdense Galaxies” (Code:
  2009L2J4MN) and by INAF/PRIN 2011: “Galaxy Evolution with the VLT
  Survey Telescope (VST)”.

  BV was supported by the World Premier International Research Center
  Initiative (WPI), MEXT, Japan and by the Kakenhi Grant-in-Aid for
  Young Scientists (B)(26870140) from the Japan Society for the
  Promotion of Science (JSPS).}

\appendix
\section{Testing the effects of the limited cluster area coverage}\label{App1}

The following three tests heve been devised with the aim of
investigating the effects of the irregular and/or limited cluster
coverage of the WINGS imaging on the results we present about the $T$-$\Sigma$
relation in Section~\ref{Sec6a} and, marginally, about the $T$-$R$ in Sec.~\ref{Sec6b}.

\subsection{Irregular Image Shape}\label{App1a}

In the first test, using the same galaxy sample of Sec.~\ref{Sec6}, we
produce a new versions of the $T$-$\Sigma$ relation in which each galaxy, with its
proper clustercentric distance ($R_B$), is weighted according to the
inverse of its circumferential coverage, that is the fraction of
circumference (of radius $R_B$) covered by the image. This is equivalent
to assume that, for each galaxy falling into the image, other
(similar) galaxies could exist at the same clustercentric distance,
but falling outside the image. This weighting also (obviously)
increases the 'nominal' number of galaxies (in particular spirals),
part of them being actually virtual objects. This test tries to
partially account for the non circular (and irregular) shape of our
images (from INT, in particular).

\begin{figure*}
\vspace{-0.5truecm}
\hspace{-0.8truecm}
\includegraphics[width=91mm]{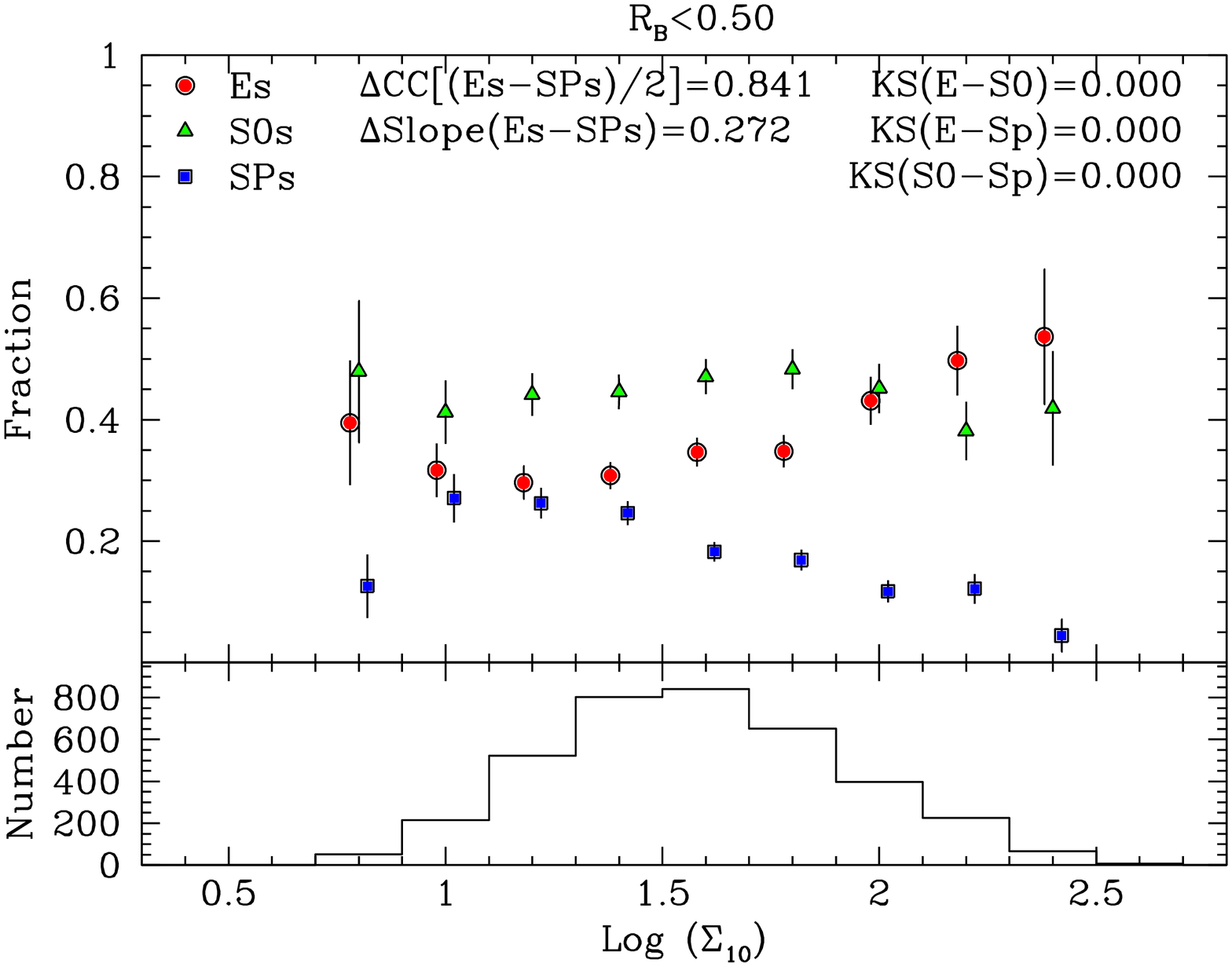}
\includegraphics[width=91mm]{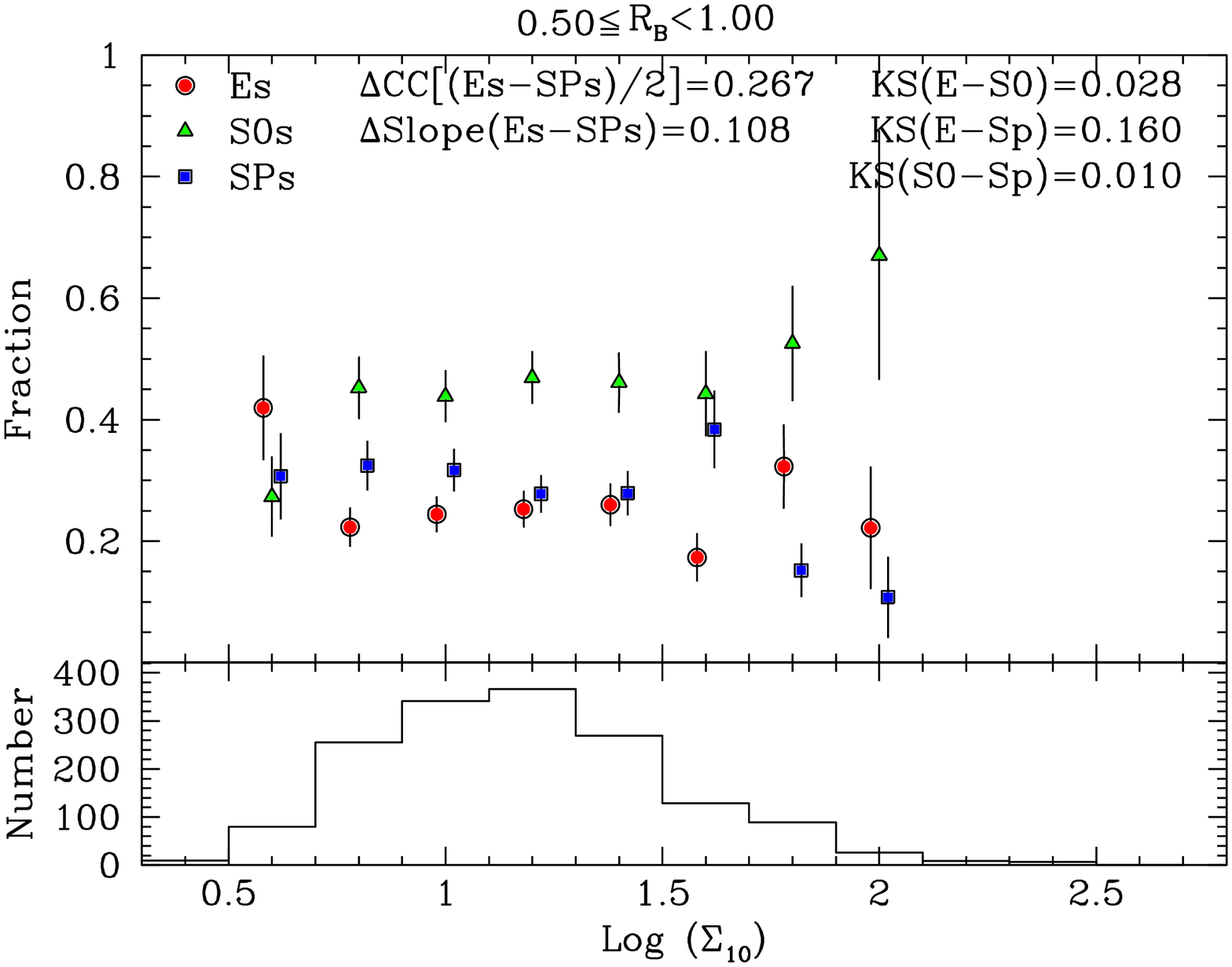}
\hspace{-0.8truecm}
\caption{$T$-$\Sigma$ for WINGS galaxies with $R_B<$0.5 (in units of
  R$_{200}$; left panel) and 0.5$\leq R_B<$1 (right panel). The
  galaxies have been weighted according to the inverse of their
  circumferential coverage (see text). The meaning of the symbols is as in
  Fig.~\ref{fig5}.}
\label{figA1}
\end{figure*}

Figure~\ref{figA1} is similar to Fig.~\ref{fig6}, but it is obtained
using the above outlined weighting procedure of galaxies. From the
comparison between figures \ref{fig6} and \ref{figA1} we conclude that
the non circular and irregular shape of the images should not
invalidate the conclusion reported in Sec.~\ref{Sec6b} that the
strenght of the $T$-$\Sigma$ relation in clusters strongly depends on the clustercentric
distance.

\subsection{Limited cluster area coverage}\label{App1b}

In the second test, we compare the $T$-$\Sigma$ obtained using galaxies in the
whole cluster sample with the corresponding ones in which only
galaxies in clusters with large values of the cluster area coverage
are included. This is equivalent to bias the analysis towards smaller
and more distant clusters.

\begin{table}
  \caption{Similar to Table~\ref{tabres}, but for just $T$-$\Sigma$ and for two different values of the minimum coverage fraction (CF)}
\begin{tabular}{ccccc}
\hline
 & \multicolumn{4}{c}{$T$-$\Sigma$} \\
\hline
CF &\ \ $R_B$ &  $\Delta$CC & $\Delta$Slope & KS(E-Sp) \\
\hline
  & 0--1 & 0.853 & 0.352 & 0.000 \\
  & $^{(1710)}$&$^{(0.065)}$&$^{(0.057)}$ & \\
 CF=0.5 & 0--0.5 & 0.878 & 0.282 & 0.000 \\
  & $^{(3351)}$&$^{(0.056)}$&$^{(0.041)}$ & \\
  & 0.5--1 & -0.087 & 0.022 & 0.210 \\
  & $^{(454)}$&$^{(0.278)}$&$^{(0.123)}$& \\
\hline
 & 0--1 & 0.782 & 0.512 & 0.000 \\
  & $^{(381)}$&$^{(0.106)}$&$^{(0.137)}$ & \\
CF=0.75 & 0--0.5 & 0.885 & 0.286 & 0.000 \\
  & $^{(2551)}$&$^{(0.051)}$&$^{(0.040)}$ & \\
  & 0.5--1 & -0.327 & -0.138 & 0.037 \\
  & $^{(127)}$&$^{(0.290)}$&$^{(0.394)}$ & \\
\hline
\end{tabular}
\label{tabcf}
\end{table}

Table~\ref{tabcf} is similar to Table~\ref{tabres}, but it reports the
results for just the $T$-$\Sigma$ and for two different values of the minimum
allowed coverage fraction (CF$>$0.5 and CF$>$0.75).  More precisely,
to assume for instance CF$>$0.5, means that just galaxies in clusters
whose CF in the considered range of $R_B$ is greater than 0.5 are each
time included in the sample. This obviously reduces the number of
employed galaxies, but makes the $T$-$\Sigma$ relation more robust against the CF issue.
From the comparison between Tables~\ref{tabres} and \ref{tabcf} we
conclude again that the limited cluster area coverage of the images
does not invalidate the conclusion reported in Sec.~\ref{Sec6b} that
the strenght of the $T$-$\Sigma$ relation in clusters strongly depends on the
clustercentric distance.

\subsection{Numerical Simulations}\label{App1c}

\begin{figure}
\vspace{-0.5truecm}
\hspace{-1.5truecm}
\includegraphics[width=100mm]{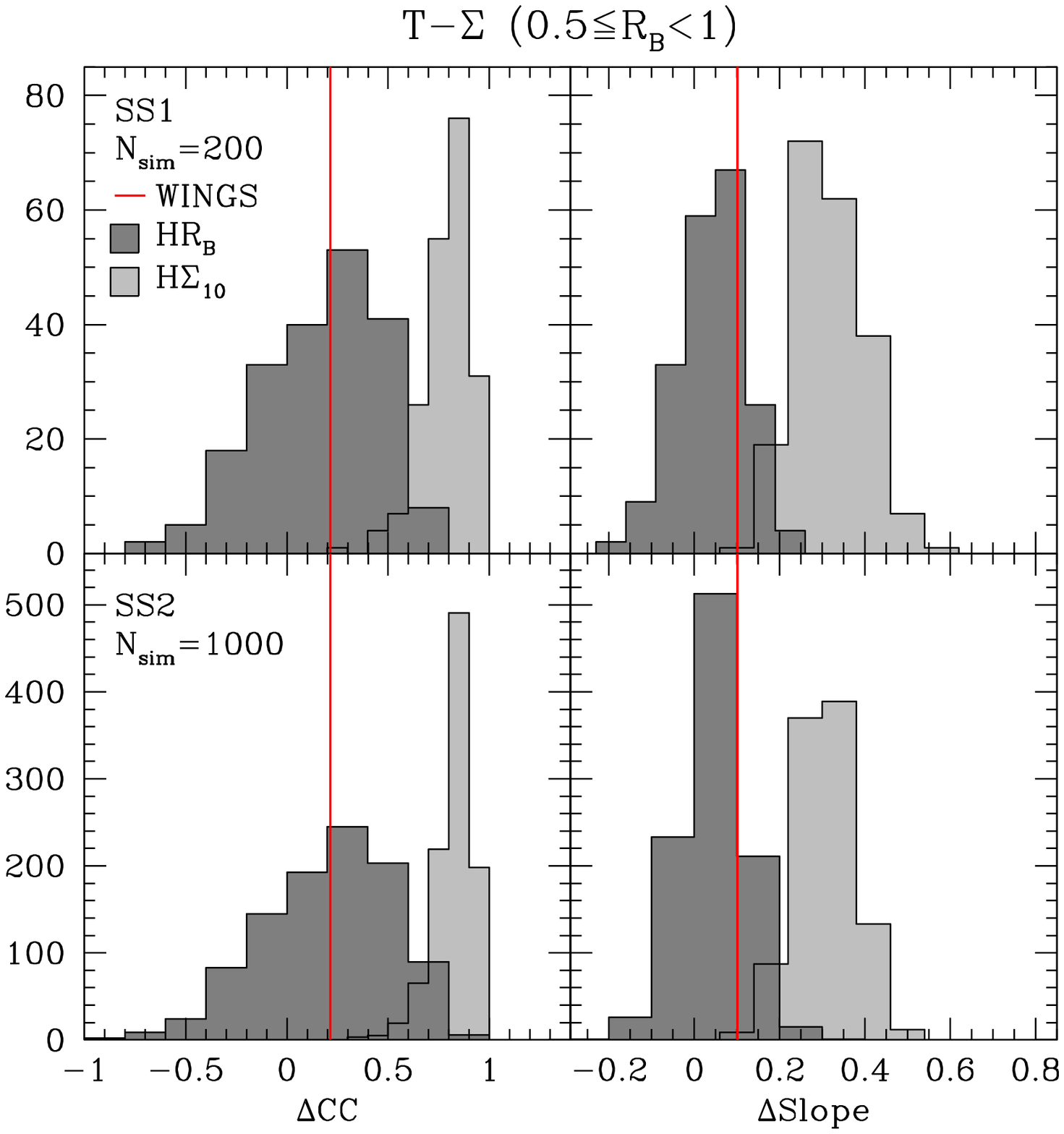}
\vspace{-3truecm}
\caption{Results of the simulations SS1 (upper panels) and SS2 (lower
  panels) for the $T$-$\Sigma$ in the $R$ interval (0.5--1). The left and right
  panels illustrate the distributions of $\Delta$CC and $\Delta$Slope,
  respectively.  The dark and light histograms refer to the H$R$ and
  H$\Sigma_{10}$ hypotheses, respectively.}
\label{figA2}
\end{figure}

Finally, we use two different sets of numerical simulations to
investigate the effect of the limited cluster area coverage on the $T$-$\Sigma$
in different ranges of the clustercentric distance. In order to
perform the simulations we have to assume some dependence of the
morphological fractions on either $R$ or $\Sigma_{10}$ (or even on
both). Therefore, these simulations can also be used, in principle, to
test which one of the two (or three) hypotheses is able to reproduce
the observed $T$-$\Sigma$ and $T$-$R$ relations.

The first set (Simulation Set 1: SS1) totally ignores the physics of
the clusters. It simply produces artificial samples of points with
random polar angles (circular symmetry) and radial coordinates
randomly extracted from a distribution suitably chosen in order to
reproduce the $R_B$--Log$\Sigma_{10}$ relation observed in the WINGS galaxy
sample. Once the $\Sigma_{10}$ values of the simulated points have been computed
in the usual way, the artificial sample is randomly thined out
following a depletion law depending on the radial coordinate and
suitably chosen in order to reproduce the variation of the total
number density of the WINGS sample as a function of $R_B$. This
procedure should properly account for the incompleteness of the real
sample due to both the irregular image shape and the limited coverage
fraction. However, the lack of physics, in particular of gravitational
clustering (2-point or 3-point correlation function), makes the
scatter of $\Sigma_{10}$ in the simulated $R$--Log$\Sigma_{10}$ relations lower than in the
real sample of WINGS galaxies. We compensate this lack of physics
through an additional scatter on $\Sigma_{10}$, tuned on the observed $R$--Log$\Sigma_{10}$
relation and depending on the radial coordinate. Then, we consider the
two alternative hypothesis that the morphological fractions ($F_E$ and
$F_S$) depend either just on $\Sigma_{10}$ (Hypothesis $\Sigma_{10}$: H$\Sigma_{10}$) or just on $R$
(H$R$). We assume in the two cases the probability that the points in
our simulated samples were elliptical or spiral galaxies to coincide
with the weighted linear fits of the relations in Fig.~\ref{fig5}
[$F_E$($\Sigma_{10}$) and $F_S$($\Sigma_{10}$)] or in Fig.~\ref{fig7} [$F_E$($R_B$) and
$F_S$($R_B$)], respectively. They are:

$$F_E=0.125+0.136\ Log\Sigma_{10};\ \ F_S=0.493-0.177\ Log\Sigma_{10} $$
$$F_E=0.401-0.25\ R_B;\ \ F_S=0.091+0.368\ R_B$$

We also formally contemplate the 'mixed' hypothesis (HMIX), in which
the morphological fractions depend on both $\Sigma_{10}$ and $R$. In this case we
roughly assume:

$$F_E=[F_E(\Sigma_{10})+F_E(R)]/2;\ \ F_S=[F_S(\Sigma_{10})+F_S(R)]/2.$$

In the second set of simulations (SS2) we adopt the same procedure to
randomly assign the morphological type, but we use the real sample of
WINGS galaxies to take the (measured) values of $\Sigma_{10}$ and $R$.

For both kinds of simulations (SS1 and SS2) and for each hypothesis
about the dependence of the morphological fractions (H$\Sigma_{10}$, H$R$, HMIX),
we make several throws, each throw producing $T$-$\Sigma$ and $T$-$R$ relations
similar to those illustrated in
Figures~\ref{fig5},\ref{fig6},\ref{fig7},\ref{fig8} ($R$ intervals:
0--1, 0--0.5, 0.5--1; Log$\Sigma_{10}$ intervals: 0--3, 0--1.45, 1.45--3), with
the same number of points (galaxies). Moreover, for each throw we
record the values of the coefficients $\Delta$CC and $\Delta$Slope
defined in Sec.~\ref{Sec6a}. Finally, in each case, we compare the
distributions of $\Delta$CC and $\Delta$Slope with the corresponding
values obtained for the real sample of WINGS galaxies (see the
previously mentioned figures). In this way, for both SS1 and SS2, we
can associate to each hypothesis (H$\Sigma_{10}$, H$R$, HMIX) and for both the $T$-$\Sigma$
and $T$-$R$ relations (in the corresponding intervals of $R$ and $\Sigma_{10}$, respectively) a
set of deviates of the observed coefficients $\Delta$CC and
$\Delta$Slope (CCdev and SLdev; both in r.m.s. units) with respect to
the corresponding distributions, as well as a set of probabilities
(P$_{CC}$ and P$_{SL}$) that these coefficients are extracted from the
distributions themselves.

Tables~\ref{tabSS1} and \ref{tabSS2} illustrate the results obtained
for the two sets of simulations (SS1 and SS2, respectively).  For each
hypothesis (H$\Sigma_{10}$, H$R$, HMIX), the tables report the deviates and the
probabilities of the coefficients $\Delta$CC and $\Delta$Slope of the
observed $T$-$\Sigma$ and $T$-$R$ relations for different intervals of $R$ and
Log$\Sigma_{10}$, respectively. They are computed comparing each coefficient
with the corresponding, proper distribution obtained from the
simulations. It should be noted that, due to the rather cumbersome
(and CPU time consuming) procedure needed to produce a single SS1
sample , the number of throws used to obtain the distributions of
coefficients CCdev, P$_{CC}$, SLdev and P$_{SL}$ is almost one order
of magnitude smaller for SS1 (N$_{sim}$=100$\div$200) than for SS2
(N$_{sim}$=1000). This fact, together with the use of the real WINGS
quantities ($\Sigma_{10}$ and $R_B$) in SS2, suggests that the robustness of the
results is greater for SS2 than for SS1. In both Table~\ref{tabSS1}
and Table~\ref{tabSS2} we report in boldface those values of CCdev and
SLdev larger than 3, as well as those values of P$_{CC}$ and P$_{SL}$
lower that 0.05. Moreover, we mark with one or two asterisks those
values of P$_{CC}$ and P$_{SL}$ equal to or less than 0.01,
respectively.

In spite of the quite rough procedure used to generate the simulated
samples in the SS1 and to attribute the morphological type to each
simulated point (in both SS1 and SS2), Tables~\ref{tabSS1} and
\ref{tabSS2} provide us with some indication about the driving
parameter of the morphological fractions in clusters:

{\it (i)} the 'canonical' hyphotesis that the morphological fractions
in clusters just depend on the local density (H$\Sigma_{10}$) is not able to
reproduce the observed $T$-$R$ relation, either for the whole sample, and for the
two $\Sigma_{10}$ intervals. This is particularly evident in Table~\ref{tabSS2};

{\it (ii)} although specifically built to obey the observed
morphology-density relation, the H$\Sigma_{10}$ simulations just
marginally reproduce the very $T$-$\Sigma$ relation in the outer part
of clusters ($R$ interval: 0.5--1), usually producing much steeper and
more correlated distributions than the observed one (see light
histograms in Figure~\ref{figA2}). While for SS1 this might be thougth
as a consequence of the quite regular shape of the simulated clusters
(see Section~\ref{Sec6c2}), such an explanation is ruled out by SS2,
where the observed WINGS quantities ($\Sigma_{10}$ and $R_B$) have
been used;

{\it (iii)} on the contrary, the alternative hyphotesis that the
morphological fractions in clusters just depend on the clustercentric
distance (H$R$) quite well reproduces both the $T$-$R$ relation (in any $\Sigma_{10}$
interval) and the $T$-$\Sigma$ relations in the outer cluster regions (see dark
histograms in Figure~\ref{figA2}), but it seems to fail in reproducing
the $T$-$\Sigma$ in the inner cluster regions. This is particularly true for
the SS1 (Table~\ref{tabSS1});

{\it (iv)} the mixed hypothesis that the morphological fractions
actually depend on both $\Sigma_{10}$ and $R$ (HMIX) turns out to be fairly
consistent with all observed $T$-$\Sigma$ and $T$-$R$ relations, but (perhaps) the
$T$-$R$ in the full Log$\Sigma_{10}$ interval (0--3).

The above remarks seem to favour the H$R$ hypothesis with respect to
the classical H$\Sigma_{10}$ scenario, although this result might critically
depend on the simulation procedure and might be modified, for
instance, by a different (not linear) representation of the relations
$F_E$($\Sigma_{10}$), $F_S$($\Sigma_{10}$), $F_E$($R$) and $F_S$($R$).

\begin{table*}
  \caption{Simulation Set 1 (SS1). Since the simulations of this set
    turn out to be rather consuming in terms of CPU time, the 
    number of throws used here to obtain the distributions of 
    coefficients CCdev, P$_{CC}$, SLdev and P$_{SL}$ varies from 100 
    to 200.}
\begin{tabular}{cllllcllll}
\multicolumn{5}{c}{$T$-$\Sigma$} & \multicolumn{5}{c}{$T$-$R$} \\
\hline
\hline
\multicolumn{10}{c}{H$\Sigma_{10}$} \\
\hline
 \ \ $R_B$ interval\ \ &  CCdev & P$_{CC}$ & SLdev & P$_{SL}$ &
 \ \ \ \ \ \ \ Log$\Sigma_{10}$ interval\ \ &  CCdev &  P$_{CC}$ & SLdev & P$_{SL}$ \\
\hline
 \ \ 0 -- 1\ \ &  1.13 & 0.32 & 0.10 & 0.90 &\ \ \ \ \ \ \ 0 -- 3\ \ & 1.10 &
 0.14 & {\bf 4.02} & {\bf 0.00**} \\
 \ \ 0 -- 0.5\ \ &  1.50 & 0.24 & 0.61 & 0.49 &\ \ \ \ \ \ \ 0 -- 1.45\ \ & 1.42 &
 0.08 & 1.90 & 0.07 \\
 \ \ 0.5 -- 1\ \ &  {\bf 4.04} & {\bf 0.01*} & 1.85 & {\bf 0.03} &\ \ \ \ \ \ \ 1.45 -- 3\ \ & 1.11 &
 0.12 & 1.82 & {\bf 0.02} \\
\hline
\hline
\multicolumn{10}{c}{H$R$} \\
\hline
 \ \ $R_B$ interval\ \ &  CCdev & P$_{CC}$ & SLdev & P$_{SL}$ &
 \ \ \ \ \ \ \ Log$\Sigma_{10}$ interval\ \ &  CCdev &  P$_{CC}$ & SLdev & P$_{SL}$ \\
\hline
 \ \ 0 -- 1\ \ &  0.11 & 0.86 & 1.79 & {\bf 0.02} &\ \ \ \ \ \ \ 0 -- 3\ \ & 1.18 &
 0.22 & 0.04 & 0.92 \\
 \ \ 0 -- 0.5\ \ &  0.83 & 0.23 & {\bf 3.21} & {\bf 0.00**} &\ \ \ \ \ \ \ 0 -- 1.45\ \ & 0.12 &
 0.92 & 0.87 & 0.34 \\
 \ \ 0.5 -- 1\ \ &  0.00 & 0.99 & 0.50 & 0.65 &\ \ \ \ \ \ \ 1.45 -- 3\ \ & 0.17 &
 0.82 & 0.07 & 0.94 \\
\hline
\hline
\multicolumn{10}{c}{HMIX} \\
\hline
 \ \ $R_B$ interval\ \ &  CCdev & P$_{CC}$ & SLdev & P$_{SL}$ &
 \ \ \ \ \ \ \ Log$\Sigma_{10}$ interval\ \ &  CCdev &  P$_{CC}$ & SLdev & P$_{SL}$ \\
\hline
 \ \ 0 -- 1\ \ &  0.10 & 0.96 & 0.93 & 0.26 &\ \ \ \ \ \ \ 0 -- 3\ \ & 0.29 &
 0.66 & 1.56 & 0.10 \\
 \ \ 0 -- 0.5\ \ &  0.14 & 0.93 & 0.99 & 0.33 &\ \ \ \ \ \ \ 0 -- 1.45\ \ & 0.70 &
 0.32 & 0.47 & 0.57 \\
 \ \ 0.5 -- 1\ \ &  1.44 & 0.31 & 0.61 & 0.60 &\ \ \ \ \ \ \ 1.45 -- 3\ \ & 0.69 &
 0.36 & 1.29 & 0.25 \\
\hline
\end{tabular}
\label{tabSS1}
\end{table*}

\begin{table*}
  \caption{Simulation Set 2 (SS2). The simulations of this set are much
    faster than in the case of SS1. Therefore, the 
    number of throws used here to obtain the distributions of 
    coefficients CCdev, P$_{CC}$, SLdev and P$_{SL}$ is everywhere
    1000.} 
\begin{tabular}{cllllcllll}
\multicolumn{5}{c}{$T$-$\Sigma$} & \multicolumn{5}{c}{$T$-$R$} \\
\hline
\hline
\multicolumn{10}{c}{H$\Sigma_{10}$} \\
\hline
 \ \ $R_B$ interval\ \ &  CCdev & P$_{CC}$ & SLdev & P$_{SL}$ &
 \ \ \ \ \ \ \ Log$\Sigma_{10}$ interval\ \ &  CCdev &  P$_{CC}$ & SLdev & P$_{SL}$ \\
\hline
 \ \ 0 -- 1\ \ &  1.11 & 0.33 & 0.19 & 0.86 &\ \ \ \ \ \ \ 0 -- 3\ \ & 1.14 &
 0.18 & {\bf 4.31} & {\bf 0.00**} \\
 \ \ 0 -- 0.5\ \ &  1.47 & 0.24 & 0.51 & 0.61 &\ \ \ \ \ \ \ 0 -- 1.45\ \ & 1.33 &
 0.07 & {\bf 3.24} & {\bf 0.00**} \\
 \ \ 0.5 -- 1\ \ &  {\bf 5.64} & {\bf 0.00**} & 2.16 & {\bf 0.04} &\ \ \ \ \ \ \ 1.45 -- 3\ \ & 1.21 &
 0.11 & {\bf 3.23} & {\bf 0.00**} \\
\hline
\hline
\multicolumn{10}{c}{H$R$} \\
\hline
 \ \ $R_B$ interval\ \ &  CCdev & P$_{CC}$ & SLdev & P$_{SL}$ &
 \ \ \ \ \ \ \ Log$\Sigma_{10}$ interval\ \ &  CCdev &  P$_{CC}$ & SLdev & P$_{SL}$ \\
\hline
 \ \ 0 -- 1\ \ &  0.28 & 0.76 & 2.29 & {\bf 0.02} &\ \ \ \ \ \ \ 0 -- 3\ \ & 1.72 &
 0.12 & 0.05 & 0.94 \\
 \ \ 0 -- 0.5\ \ &  0.66 & 0.44 & 2.72 & {\bf 0.01*} &\ \ \ \ \ \ \ 0 -- 1.45\ \ & 1.72 &
 0.18 & 1.53 & 0.14 \\
 \ \ 0.5 -- 1\ \ &  0.03 & 0.97 & 0.51 & 0.61 &\ \ \ \ \ \ \ 1.45 -- 3\ \ & 0.98 &
 0.41 & 0.13 & 0.92 \\
\hline
\hline
\multicolumn{10}{c}{HMIX} \\
\hline
 \ \ $R_B$ interval\ \ &  CCdev & P$_{CC}$ & SLdev & P$_{SL}$ &
 \ \ \ \ \ \ \ Log$\Sigma_{10}$ interval\ \ &  CCdev &  P$_{CC}$ & SLdev & P$_{SL}$ \\
\hline
 \ \ 0 -- 1\ \ &  0.30 & 0.77 & 1.25 & 0.18 &\ \ \ \ \ \ \ 0 -- 3\ \ & 0.00 &
 0.99 & 2.21 & {\bf 0.03} \\
 \ \ 0 -- 0.5\ \ &  0.21 & 0.82 & 1.19 & 0.26 &\ \ \ \ \ \ \ 0 -- 1.45\ \ & 0.34 &
 0.73 & 0.84 & 0.41 \\
 \ \ 0.5 -- 1\ \ &  1.76 & 0.18 & 0.74 & 0.44 &\ \ \ \ \ \ \ 1.45 -- 3\ \ & 0.48 &
 0.61 & 1.59 & 0.10 \\
\hline
\end{tabular}
\label{tabSS2}
\end{table*}

\end{document}